\RequirePackage{amsthm}

\documentclass[pdflatex,sn-basic,Numbered]{sn-jnl}

\usepackage{graphicx}%
\usepackage{multirow}%
\usepackage{amsmath,amssymb,amsfonts}%
\usepackage{amsthm}%
\usepackage{mathrsfs}%
\usepackage[title]{appendix}%
\usepackage{xcolor}%
\usepackage{textcomp}%
\usepackage{manyfoot}%
\usepackage{booktabs}%
\usepackage{algorithm}%
\usepackage{algorithmicx}%
\usepackage{algpseudocode}%
\usepackage{listings}%
\usepackage[numbers]{natbib}
\usepackage{verbatim}
\usepackage{amsmath}
\usepackage{bbm}
\usepackage{tabularx}
\usepackage{makecell}
\usepackage{ragged2e}
\usepackage{enumitem} 
\usepackage{adjustbox}
\usepackage{float}
\usepackage{lmodern}
\usepackage{epstopdf}
\usepackage{multirow}
\usepackage{caption}
\usepackage{subcaption}

\theoremstyle{thmstyleone}%
\newtheorem{theorem}{Theorem}
\newtheorem{proposition}[theorem]{Proposition}%

\theoremstyle{thmstyletwo}%

\theoremstyle{thmstylethree}%
\newtheorem{definition}{Definition}%

\newcommand\restr[2]{{
  \left.\kern-\nulldelimiterspace 
  #1 
  \littletaller 
  \right|_{#2} 
  }}

\newcommand{\R}{\mathbb{R}}

\newcommand{\Rv}{{\pmb{R}}}
\newcommand{\A}{\mathcal{A}}
\newcommand{\T}{\mathcal{T}}

\newcommand{\abs}[1]{{\left|#1\right|}}
\DeclareMathOperator*{\argmax}{arg\,max}
\DeclareMathOperator*{\argmin}{arg\,min}

\newcolumntype{Y}{>{\Centering\arraybackslash}X}

\begin{document}

\title[Article Title]{Learning the Value Systems of Agents with Preference-based and Inverse Reinforcement Learning}

\author*[1]{\fnm{Andr{\'e}s} \sur{Holgado-S{\'a}nchez} (ORCID: {0000-0001-8853-1022})}\email{andres.holgado@urjc.es}

\author[1]{\fnm{Holger} \sur{Billhardt} (ORCID: 0000-0001-8298-4178)}\email{holger.billhardt@urjc.es}
\equalcont{These authors contributed equally to this work.}

\author[1]{\fnm{Alberto} \sur{Fern{\'a}ndez} (ORCID: 0000-0002-8962-6856)}\email{alberto.fernandez@urjc.es}
\equalcont{These authors contributed equally to this work.}

\author[1]{\fnm{Sascha} \sur{Ossowski} (ORCID: {0000-0003-2483-9508})}\email{sascha.ossowski@urjc.es}
\equalcont{These authors contributed equally to this work.}

\affil*[1]{\orgdiv{CETINIA}, \orgname{Universidad Rey Juan Carlos}, \orgaddress{\street{Tulip{\'a}n (Unnumbered)}, \city{M{\'o}stoles}, \postcode{28933}, \state{Madrid}, \country{Spain}}}

\abstract{
Agreement Technologies refer to open computer systems in which autonomous software agents interact with one another, typically on behalf of humans, in order to come to mutually acceptable agreements. With the advance of AI systems in recent years, it has become apparent that such agreements, in order to be acceptable to the involved parties, must remain aligned with ethical principles and moral values. However, this is notoriously difficult to ensure, especially as different human users (and their software agents) may hold different value systems, i.e. they may differently weigh the importance of individual moral values. Furthermore, it is often hard to specify the precise meaning of a value in a particular context in a computational manner.
Methods to estimate value systems based on human-engineered specifications, e.g. based on value surveys, are limited in scale due to the need for intense human moderation. In this article, we propose a novel method to automatically \emph{learn} value systems from observations and human demonstrations. In particular, we propose a formal model of the \textit{value system learning} problem, its instantiation to sequential decision-making domains based on multi-objective Markov decision processes, as well as tailored preference-based and inverse reinforcement learning algorithms to infer value grounding functions and value systems. The approach is illustrated and evaluated by two simulated use cases.}




\maketitle
\section*{Ethics declarations}
\textbf{Conflict of interest}. The authors have no competing interests to declare that are relevant to the content of this article.

\section*{Acknowledgements}

\textit{This version of the article has been accepted for publication, after peer review but is
not the Version of Record and does not reflect post-acceptance improvements, or any corrections. The Version of Record is
available online at: \url{http://dx.doi.org/10.1007/s10458-026-09732-0}}.

This work has been supported by grant VAE: TED2021-131295B-C33 funded by MCIN/AEI/10.13039/501100011033 and by the “European Union NextGenerationEU/PRTR”, by grant COSASS: PID2021-123673OB-C32 funded by MCIN/AEI/10.13039/501100011033 and by “ERDF A way of making Europe”, and by project grant EVASAI: PID2024-158227NB-C32 funded by MICIU/AEI/10.13039/501100011033/FEDER, UE. Andr{\'e}s Holgado-S{\'a}nchez has received funding  
by grant ``Contratos Predoctorales de Personal Investigador en
Formaci{\'o}n en Departamentos de la Universidad Rey Juan Carlos (C1 PREDOC 2025)'', funded by Universidad Rey Juan Carlos.

\section{Introduction}\label{sec1}
Agreement Technologies (AT) refer to a sandbox of methods and tools to enable and to govern the interactions of intelligent software agents, so as to shape their autonomy and help them reach agreements that are acceptable to all stakeholders. In open multiagent systems, and AI systems in general, autonomy should be exercised in a responsible manner~\cite{virginiaDignumResponsibleAutonomy2017} so that  decisions and behaviour obey ethical principles and human values. This requirement has been referred to as  \textit{value alignment}~\cite{Russell2022alignmentDefinition,Gabriel2020alignment} of (distributed) AI systems. In recent years, efforts to engineer value-aligned AI systems led to proposals implementing the notion of \emph{value-awareness} in autonomous agents~\cite{values2021nardineold,valueengineeringAutonomous2023}, where agents are endowed with explicit representations of human values and the ability to reason with and about them.

Once there is an agreement on the meaning of a particular value in a specific domain, and a computational way to express the level of alignment of entities with that value (either quantitatively or in terms of qualitative preferences), different types of reasoning can be performed in systems of value-aware agents. Several proposals have been put forward that dynamically achieve value alignment by defining value-based agent decision-making strategies~\cite{manel2022ethical,andres2023eumas,andres2023vae,andres2024vecompPaper,rodriguez2026reinforcementEthicalEmbedingWeightsRewardRL}, selecting different system configurations (i.e. norms)~\cite{montes2022synthesis,serramia2023EncodingValueAlignedNorms},  or aggregating value-based preferences (sometimes called \textit{value systems})~\cite{leraleri2022aggregation,leraleri2024aggregation,Liscio2023ValuesInSocioTechicalSystemsValueInference}. However, such computational representations of values are notoriously difficult to engineer. 

Russell~\cite{Russell2022alignmentDefinition} synthesises three ambitious principles for designing truly generally value-aligned machines. They should not only (1) maximize the realization of human preferences, and (2) know how the preferences are defined, but (3) should also be able to learn them from human behaviour. Some recent works follow this idea~\cite{CooperativeIRLHadfield2016,leike2020} by mimicking agent behaviour using different forms of imitation. However, such approaches are often limited by the fact that learned behaviours cannot be explained in terms of the underlying ethical principles.

In this paper, we are interested in the problem of learning explicit specifications of values from human demonstrations, a problem that has been labelled as \textit{value learning}~\cite{Soares2018ValueLearningProblem,Sezener2015inferringAGI}. In particular, setting out from a previously identified set of labels for the values that are relevant in a particular domain, we are interested in learning  both, a specific computational representation of the values understood as a measure of value alignment over world entities (the value \textit{grounding function}), as well as  an explicit expression of the particular and possibly heterogeneous value preferences of the agents (their \textit{value systems}). To address this \textit{value system learning problem}, we revisit and enhance the value system representation model and problem solution framework proposed in previous work~\cite{andres2024vecompPaper}. 


In particular, we propose a solution to the value system learning problem for value-based agents, whose behaviour is primarily guided by values, and that face a sequential decision problem. The model of the latter sets out from Markov Decision Processes (MDP) while the former, in line with \cite{manel2022ethical,montes2022synthesis,yangRewardsInContextMOAlignment,wang-etal-2024-interpretableRewardModelingLLM}, takes a  consequentialist stance on individual values and value systems. More specifically, we formalize the value system learning problem in terms of multi-objective Markov Decision Processes (MOMDPs), by encoding separate values as different components of a reward vector. This allows an AI system to recognise these values and align with them under varying conditions~\cite{Vamplew2018Human-alignedProblem}. Additionally, we model the value system of each particular agent through a weighted linear scalarization of the reward components. On the basis of this representation, we approach the proposed learning problem using adaptations of existing algorithms for both classical Inverse Reinforcement Learning~\cite{surveyIRLArora2021} 
and for Preference-based (inverse) Reinforcement Learning~\cite{surveyRLFromPreferencesPbRL,trexpreferences2019}. 

We evaluate the proposed framework and algorithms in two simulated use cases, modelled as MOMDPs with discrete state and action spaces. The first use case encompasses the simulated firefighter scenario proposed by~\cite{osman2025instillingorganisationalvaluesfirefighters}, where learning the grounding of the values is specially challenging. The second use case refers to value-based route choice~\cite{routechoicemodelingPRATO2009} in a scenario inspired by the Shanghai road network~\cite{zhao2023routesairl}, where learning an agent's value system is complicated by the correlations among the different values. 

The paper is structured as follows. Section~\ref{sec:background} provides background on the fields of agreement technologies and value-awareness engineering, and reviews related work on value learning as well as algorithms for inverse reinforcement learning and preference-based reinforcement learning. In Section~\ref{sec:values-framework} we explain our proposed value system learning framework and the problem adaptation to sequential decision-making scenarios. Section~\ref{sec:value-system-learning-algorithms} presents specific algorithms for solving the value system learning problem within our MOMDP formulation. In Section~\ref{sec:eval} we evaluate our proposal in the Firefighters and Roadworld use cases and provide a discussion of the results.  Section~\ref{sec:conclusion} describes the lessons learned from our work, discusses its limitations, and points to future lines of work. 

\section{Background}\label{sec:background}

In this section we discuss the state of the art in different fields that are relevant to our work. We set out from a quick overview of the field of Agreement Technologies and relate it to the notion of responsible autonomy. We then discuss different approaches to model value awareness for artificial agents, especially in the context of sequential decisions. Section \ref{sec:background-learningvalues} summarises existing work on learning value specifications from behaviour demonstrations. Finally, we point to different works on inverse and preference-based reinforcement learning that are relevant to the value-system learning approach presented in this article.

\subsection{Agreement Technologies}\label{sec:background-agreement}

The term ``Agreement Technologies'' was first used by both Michael Wooldridge and Nicholas R. Jennings in the early two-thousands, referring to the integration of several techniques used in the field of multiagent systems.   At that time,  large-scale open distributed systems arose as an area of significant social and economic potential, and peer-to-peer based interaction schemes, as the ones developed by the multiagent systems community, were considered fundamental to achieve that goal. The intuitions evoked by the term were further grounded and related to each other by Carles Sierra in the context of the Spanish Consolider Project Agreement Technologies (AT), and later on within the European COST Action on AT.  The vision of next-generation open distributed systems based on the concept of agreement between computational agents was put forward, with a normative context defining the ``rules of the game'' and a ``call-by agreement'' interaction method supporting both the establishment and the enactment of agreements. 

This proposal was further structured through an architecture sketch that later became known as ``AT layer cake'' or ``AT tower''~\cite{Springer-AT}. \emph{Semantic technologies} would provide solutions to semantic mismatches through the alignment of ontologies, so agents could reach a common understanding of the elements of agreements. In this manner, a shared multi-faceted ``space'' of agreements would emerge, providing essential information to the remaining layers. The next level was concerned with the definition of \emph{norms} determining constraints that the agreements, and the processes leading to them, should satisfy. Thus, norms would be conceived of as a means of ``shaping'' the space of valid agreements. \emph{Organisations} would further restrict the way agreements are reached by imposing organisational structures on the agents. They would thus constitute a way to efficiently design and evolve the space of valid agreements, possibly based on normative concepts. The \emph{argumentation and negotiation} layer provided methods for reaching agreements that respected the constraints that norms and organisations would impose on the agents. This could be seen as choosing certain points in the space of valid agreements. Finally, the \emph{trust and reputation} layer would keep track of whether the agreements reached, and their executions, would respect the constraints put forward by norms and organisations. As such, it would complement the other techniques that shaped the ``agreement space'', by relying on social mechanisms that interpret the behaviour of agents.

A key characteristic of Agreement Technologies is that they provide a sandbox of methods to shape agent autonomy from a both \emph{social} and \emph{normative} perspective, and through an explicit and transparent agreement process. Still, it has recently been argued that computational intelligent agents should use their autonomy in a \emph{responsible} manner, i.e. AI systems should be designed to take ethical considerations into account and to consider the moral consequences of their actions and decisions, in accountable, responsible, and transparent ways~\cite{virginiaDignumResponsibleAutonomy2017}. For Agreement Technologies to play a relevant role to this respect, it is necessary to further investigate how to ensure that the agreement reached by autonomous agents respect ethical principles and are aligned with moral values. A promising approach to these problems is to advance towards the engineering of \emph{value-aware} agents~\cite{values2021nardineold,valueengineeringAutonomous2023}, endowed with explicit representations of human values and the ability to reason with and about them.

\subsection{Value awareness and ethical decision-making}\label{sec:background-new}

One of the questions that value-awareness proposals face is what are values and which values are important for AI systems to have awareness of. Different \textit{value theories} propose a definition and a delimited set of different human values such as the Schwartz's theory of Basic Human Values (BHV)~\cite{schwartz1992universals} or the Haidt's Moral Foundations Theory~\cite{mft}. Some values are particular to specific contexts, as values (continuously) emerge in different application domains~\cite{vandePoel2021}. For this reason, value sensitive design~\cite{vsdreal2006}, argues that values are open-ended concepts that should be elicited by the stakeholders in particular domains. In general, Rohan's~\cite{Rohan2000} concerns on the misalignment of value theories proposed by psychologists and non-psychologists, motivated the synthesis of an operational definition of values shared by value-awareness works as ``abstract concepts that guide behaviour whose interpretations depends on context''~\cite{osman2024computationalNUEVO}.

Value-awareness is heavily related to the concept of Artificial Moral Agents (AMA)~\cite{wallach2008moral}, i.e. autonomous agents that are able to behave and reason in alignment with moral and societal values. Value-aware agents extend AMAs in order to deal with the plethora of values (understood as in the previous definition) that influence actual human choices in varying contexts~\cite{valueengineeringAutonomous2023}. A main goal in the field is to build autonomous systems that are capable of explicitly aligning with the \textit{pluralistic} \textit{value systems}~\cite{leraleri2024aggregation} of stakeholders, i.e. the decision-making model based on human values that governs their value-based decisions. This \textit{value-based (or value-aligned) decision making} can be viewed as a particular case of ethical deliberation~\cite{virginiaDignumResponsibleAutonomy2017} where the ethical decisions are those that align with a certain value system.

A natural question is how to design this value-based decision making process and properly implement it in an autonomous agent. Authors from moral psychology such as Haidt~\cite{haidt2001emotional}  defend that ethical judgements in humans are driven primarily by fast, automatic intuitions and feelings rather than by conscious, deliberative calculations of how different alternatives align with a person’s explicit values. However, when emotions at play collide, moral judgements often require complex reasoning~\cite{GREENE2002517}. In any case, in order to build value-aligned artificial agents we require an operationalization~\cite{Shahin2022OperationalizingvaluesSurveySoftwareengineering} of values into explicit computable constructs to permit value-aware reasoning and decision-making, even if  this implies certain simplifications.

Recognizing that reasoning abilities are required to build, in particular, AMAs (and by extension, value-aware agents), Dignum~\cite{virginiaDignumResponsibleAutonomy2017} analyses the possible implementations of the most utilized ethical theories for ethical deliberation: consequentialism, deontology and virtue ethics. Applied to model value-based decision making, the most utilized in existing proposals on value-awareness   is, by far, consequentialism, which defines the "morally right action is the one that produces a good outcome"~\cite{virginiaDignumResponsibleAutonomy2017}. In particular, most of these works follow a \textit{pluralistic} form of \textit{utilitarianism} (welfarist utilitarianism) that sees values as different perspectives to the goodness of an action~\cite{Sinnott-Armstrong2019-SINC-5}. For instance, Montes et al.~\cite{montes2022synthesis} measures the \textit{value alignment} of norms in terms of the final outcomes they permit through a \textit{semantics function} that operationalizes (or  \textit{grounds}~\cite{andres2024vecompPaper}) the meaning of a value to a particular domain.

Some authors try to represent the \textit{value systems} of different stakeholders, via a weighting or ordering scheme over the pluralistic values that governs the stakeholders' value-based decisions~\cite{Beauchamp2019valuesystemasorderedvalues,bench2012using,serramia2023EncodingValueAlignedNorms,rodriguez2026reinforcementEthicalEmbedingWeightsRewardRL}. Prominent examples are the proposals by Serramia et al.~\cite{Serramia2018,serramia2023EncodingValueAlignedNorms} on selecting a set of norms (stated as deontic rules) that are the most aligned with given value system, expressed by a set of ordered values. Rodr{\'i}guez-Soto et al.~\cite{manel2022ethical,rodriguez2026reinforcementEthicalEmbedingWeightsRewardRL} present a reinforcement learning agent that simultaneously abides by its value system (understood as a set of values with a total order) and respects given norms, while still pursuing its own particular goals, defining value alignment and norm enforcement via reward functions (thus, aligning directly with welfarist utilitarianism).

More recent value system representation models and frameworks, while maintaining a welfarist utilitarian view, take into account further aspects in the actual meanings of the values and their interactions. Karanik et al.~\cite{Karanik2024} proposes a value system constructed using the order of importance established by the agent on the values (quantitatively) and with a theory of values determining the interaction between them (specifically, with interest in Schwartz's theory~\cite{schwartz1992universals}). Inspired by a careful analysis of multiple value theories, Osman and D'Inverno~\cite{osman2024computationalNUEVO} propose an alternative representation approach based on so-called ``value taxonomies'' that represent context-dependent values and their priorities through conceptual hierarchies, proposing a way to design the computation of the value alignment of different entities based on their value-related properties and their aggregation in line with the structure of the particular taxonomy considered.

\subsection{Learning value specifications from demonstrations}\label{sec:background-learningvalues}

The previous approaches for modelling value awareness face the risks of misspecification~\cite{Gabriel2020alignment,Balakrishnan2019behaviouralalignment,CooperativeIRLHadfield2016,Sumers2022InstructionsAndDescriptions}. Hence, an important research direction focuses on learning adaptable value specifications from demonstrations for different application areas. Soares~\cite{Soares2018ValueLearningProblem} refers to this broad challenge as the \textit{value learning problem}. To our knowledge, there is no proposal that addresses this problem as a whole, but different aspects of it have been studied in isolation.

Liscio et al~\cite{Liscio2023ValuesInSocioTechicalSystemsValueInference} proposed a modular framework of \textit{value inference} in socio-technical systems consisting of three steps.
Firstly,  \textit{Value identification}  consists of determining the set of values relevant to a decision context. Participatory methods used in value sensitive design~\cite{vsdreal2006}, hybrid approaches with stakeholder conversations~\cite{Liscio2023axiesContextSpecificValueIdentification}, by elicitation of preferences through experiments~\cite{googleVeilOfIgnorance2023} or with data-driven methods such as classification~\cite{QiuZhaoLiLuPengGaoZhu2022ValueNetDataset} and text processing~\cite{wilson2018valueidentificationtext} (sometimes enriched with neuro-symbolical representations of values~\cite{lazzari2024explainablemoralvaluesneurosymbolic}) have been used. 
Secondly \textit{value system estimation} is used to learn  preferences over values based on observed decisions, texts or stakeholder conversations~\cite{liscio-etal-2022-cross,Liscio2023InferringHybridValues}. 
Thirdly,  \textit{value aggregation}~\cite{leraleri2022aggregation,leraleri2024aggregation} consists of aggregating the possibly heterogeneous value systems of various agents for consensus.

 Some ethical behaviour learning techniques can be conceived  as forms of value learning. From the \textit{machine ethics} literature~\cite{Anderson2006} comes an approach called \textit{prima facie} duty learning~\cite{Anderson2018GenEthILP}, that directly learns general action preferences (principles) in terms of the ethical features they provoke or entail, from observed action preferences. Notably, this approach learns a form of deontic principles, hence, it oposses the typical utilitarian value specifications of other works. However, the learned principles may not be related to actual human values, and the action preferences are assumed to be global to all stakeholders. 
 The work reported in~\cite{userStudyLearningValues} infers alignment functions (e.g. as per~\cite{montes2022synthesis}) based on user studies for four fundamental bioethical principles (beneficence, non-maleficence, autonomy, and justice) in the healthcare domain, which can be conceived as a form of value learning from (highly structured) demonstrations.

 Authors such as Dignum~\cite{virginiaDignumResponsibleAutonomy2017}, Soares~\cite{Soares2018ValueLearningProblem} or Leike et al.~\cite{Leike2018ScalableAA} admit that a (utilitarian) reinforcement learning formulation is applicable to approach the value-aligned decision-making problem. However, they are aware of the problem of reward misspecification~\cite{Soares2018ValueLearningProblem,Sumers2022InstructionsAndDescriptions} that can lead to misaligned behaviours. The field of \textit{inverse reinforcement learning} could provide a solution to the issues that arise from manual reward engineering~\cite{Soares2018ValueLearningProblem} (see Section~\ref{sec:stateofartIRL}). A notable example of the application of this technique is CIRL (Cooperative Inverse Reinforcement Learning)~\cite{CooperativeIRLHadfield2016} where human-aligned behaviour is learned in cooperation games through imitation and teaching. Here, value alignment is implicitly learned so no computational representation of the meaning of specific values is extracted. In the same line, inferring reinforcement learning rewards from human feedback has been proposed to ensure alignment with human behaviour~\cite{Leike2018ScalableAA,leike2020,dpoLLM2023}. Still, the rewards learned are, generally, not explicitly linked to specific human values or ethical principles. 
     
\subsection{Inverse and Preference-based Reinforcement Learning }\label{sec:stateofartIRL}

The problem of Inverse Reinforcement Learning (IRL) in Markov Decision Processes (MDP) refers to extracting a reward function from observed behaviours~\cite{ng2000algorithms}. IRL assumes that the seen behaviours/demonstrations in an MDP correspond to an ``expert policy'' $\pi_E$, which is assumed to be optimal with regard to some unknown reward function $R$, i.e. the policy chooses actions at each state of the MDP so as to maximize the expected cumulative reward. The objective is to infer another reward function $\hat{R}$ so that the corresponding optimal policy for the MDP produces the same (or sufficiently similar) behaviour to that of the expert.

The IRL problem is, however, ill-posed as shown by~\cite{ng2000algorithms}: many reward functions can explain the same behaviour. Different additional desirable criteria were added to deal with this issue, like minimizing the \textit{feature expectation difference} with the expert policy~\cite{ApprenticeshipIRLAbbeel2004}, leveraging measures of  entropy~\cite{maximumEntropyIRLZiebart2008,MaximumCausalEntropyIRLZiebart2010,DeepMaximumEntropyIRLWulfmeier2015}, or maximizing the likelihood of a cost function (negative reward) being optimal in the expert demonstrations~\cite{GCLFinn2016}.
Other approaches are Inverse Action-value Iteration~\cite{DeepQIRLKalweit2020}, which solves an inverse form of Q learning; Gaussian IRL~\cite{GaussianIRLLevine2011}, which uses a Bayesian formulation; and Adversarial Inverse RL~\cite{AIRLFu2018}, that takes direct inspiration from generative adversarial networks (GANs). 

Preference-based reinforcement learning (PbRL)~\cite{surveyRLFromPreferencesPbRL,kaufmann2024survey} algorithms have been proposed that can directly learn from an expert's preferences  by observing pairwise comparisons between trajectories in an MDP (i.e. pairs of the form $\tau \succ \tau'$, where $\tau = ((s_0, a_0), \dots)$ is a sequence of states, actions, or both). Most of these works rely on fitting a reward model~\cite{christiano2023deeprlpreferences} so that a (stochastic) policy $\pi: S\times A \to [0,1]$, which maximizes that reward (under a given horizon or discount factor), will optimize the probability of selecting the best trajectories. As such, some PbRL approaches like~\cite{trexpreferences2019} can be considered an alternative to classical IRL as they learn reward functions using pairwise comparisons instead of plain trajectories.

\section{Value System Learning}\label{sec:values-framework}

This section details our approach to a value system learning framework. We begin by defining and formalizing key concepts, including values, value alignment functions, grounding functions, value systems and value system functions. We then characterise the value system learning problem in general and in the context of sequential decision-making processes.
A glossary of the terms used in the following sections is given in Table~\ref{tab:glossary-of-terms}, Appendix~\ref{sec:glossary}.

\subsection{Value system representation model and learning framework}

We first introduce the notion of \textit{value} used in our model. It should be general enough to be able to capture key aspects of common value theories (e.g. Schwartz's~\cite{schwartz1992universals} or Curry's~\cite{Curry2022}), but also sufficiently explicit to account for a computational operationalization. In particular, we adopt Osman and D'Inverno's notion of values as ``human abstract concepts that guide behaviour''~\cite{osman2024computationalNUEVO}, which is a sufficiently abstract definition that is shared among many value theories~\cite{andres2024icaart} and values that appear in new domains~\cite{vsdreal2006}. 

In our framework, we set out from a set of $m$ values $V=\{v_1,\dots,v_m\}$. Each value $v_i \in V$ constitutes essentially a label that gives a name to a value. When \textit{grounded} in a particular domain, a value label acquires a specific meaning. We model this meaning in terms of a value alignment function that allows to measure the alignment of \textit{entities} present in this domain with a given value $v_i$. {Depending on the domain, the set of entities might be the set of alternatives in a classical decision-making stance, or, rather, the outcomes that these alternatives provoke. For example, in route choice analysis, the entities of study could be the paths or routes that the agent can traverse; whereas in government policy making, the set of entities could consist of the outcomes that these policies provoke in society.

The notion of alignment constitutes a (quantitative) measure as to how much an entity promotes (or demotes) a value $v_i$. Thus, our model is fundamentally consequentialist or utilitarian, as advanced in Section~\ref{sec:background-new}. Different alternatives are not ruled by ethical imperatives nor influenced by virtues~\cite{virginiaDignumResponsibleAutonomy2017}, rather, they are evaluated at the moment of deliberation in terms of their value alignment. While this model does not necessarily correspond to the reasoning process of humans when realizing moral (and value) judgements (or its lack of it~\cite{haidt2001emotional}), we hypothesize it can serve artificial value-aware agents to represent the human value-based deliberation outcomes with a sufficient precision\footnote{We plan on evaluating the extent by which this hypothesis holds in real-world scenarios in future works.}.


\begin{definition}[Value Alignment Function]\label{def:value-alignment-function}
     Given a value $v_i$, the function $\A_{v_i}: E \to \R$ is a \textbf{value alignment function} for $v_i$. 
\end{definition}

A value alignment function $\A_{v_i}$ induces a preference relation $\preccurlyeq_{v_i}$ over entities with respect to value ${v_i}$. That is, for all $e,e' \in E$, if  $e'$ is weakly preferred over $e$ it holds that:
    $$e \preccurlyeq_{v_i} e' \iff \A_{v_i}(e) \leq \A_{v_i}(e') $$

We will illustrate our framework with a simple running example on car trip route choice. We consider two possible routes from a particular origin to a destination, $e_1$ and $e_2$. Each of those routes constitutes an entity in our domain. We assume there are two values in this domain that are of particular interest: \emph{efficiency}\footnote{Rooted in utilitarianism, we conceive efficiency as a technical value that correlates negatively with cost and positively with pleasure. For simplicity, in this example, efficiency is measured in negative time units.} ($v_1$) and \emph{sustainability} ($v_2$). These values constitute the set $V=\{v_1,v_2\}$. For the travellers, the alignment function $A_{v_1}$ with value $v_1$ (efficiency) could defined as the time costs (a negative amount) of the trip. The alignment function $A_{v_2}$ with $v_2$ (sustainability) could be defined as a cost related to the total car's fuel consumption for the trip\footnote{We assume, for simplicity, this consumption depends only on the average consumption per type of road and not on the driver's style.}. We assume that route $e_1$ is a short trip through the city centre but that is highly fuel-inefficient due to congestion that has the following alignments $\A_{v_1}(e_1) = -1.8$ and $\A_{v_2}(e_1) = -2.0$. On the other hand, route $e_2$ is a highway trip that requires slightly more time to travel but is much less fuel-intensive, thus, we could have $\A_{v_1}(e_2) = -2.1$  and $\A_{v_2}(e_2) = -1.2$. 

To specify the semantics of a set of values, we define the notion of \textit{grounding function}.

\begin{definition}[Grounding function]\label{def:grounding}
     Given a set of values $V$, we define the vector of value alignment functions for $V$ as a \textbf{grounding function} of $V$: $G_V=(\A_{v_1},\dots,\A_{v_m})$ 
\end{definition}

In our running example, the grounding function for values $V=\{v_1,v_2\}$ would be given by $G_V(e)=\left(\A_{v_1}(e),\A_{v_2}(e)\right)\in \R^2$. Thus, $G_V(e_1)=(-1.8,-2.0)$ and $G_V(e_2)=(-2.1,-1.2)$.

In our computational model, we assume that artificial agents build their individual value systems on top of a grounding function, i.e. taking into account the alignment functions of the values $v_i \in V$ that are relevant within a certain domain. 

\begin{definition}[Value system]\label{def:value-system}
Let $V = \{v_1, ..., v_m\}$ be a finite set of values, and $G_V=(\A_{v_1},..., \A_{v_m})$ be a grounding function for $V$. The \textbf{value system} of an agent $j$ derived from the grounding function $G_V$ is a weak order $\preccurlyeq^j_{G_V}$ over $E$.

\end{definition}

Notice that an agent's value system aggregates grounded values, based on the importance that different values have for that agent. 
Given a grounding function, a value system of an agent can be represented by means of a value system function.

\begin{definition}[Value System Function]\label{def:value-system-alignment-function}
Let $V$ be a set of values, and $j$ an agent with value system $\preccurlyeq^j_{G_V}$. The function $\A_{f_j,{G_V}} : E \to \R$ with $\A_{f_j,{G_V}}(e) = f_j(\A_{v_1}(e), \dots, \A_{v_m}(e))$ is a \textbf{value system function} for agent $j$ if it represents $\preccurlyeq^j_{G_V}$ over $E$, i.e. for all $e,e' \in E$:
$$\A_{f_j,{G_V}}(e) \leq \A_{f_j,{G_V}}(e') \iff e \preccurlyeq^j_{G_V} e'$$ where $f_j: {\R}^m \to \R$ is the \textit{value system aggregation function} that $j$ applies to combine the alignments of $e$ with respect to each value $v_i \in V$. 
\end{definition} 

In our running example, suppose that for a particular agent $j$, the value of efficiency is twice as important as sustainability (both measured in terms of the presented grounding $G_V$). This fact is expressed by a value system aggregation function $f_j(x,y) = 2x + y$. This leads to the value system function for agent $j$, $A_{f_{j},{G_V}}$, defined by $A_{f_j,{G_V}}(e_1) = -5.6$ and $A_{f_j,{G_V}}(e_2) = -5.4$.

Notice that in our model a value system function need not be the result of a hierarchical application of values in line with their importance for an agent $j$ (as other approaches assume~\cite{bench2012using,Serramia2018}). Indeed, according to agent $j$'s value system, route $e_2$ is preferred over $e_1$ ($e_1 \preccurlyeq^j_{G_V} e_2$), even though $e_2$ would be preferred over $e_1$ if only the ``most important'' value for $j$ (efficiency, $v_1$) was considered ($e_1 \succcurlyeq_{v_1} e_2$). This example also shows that value system aggregation functions must be interpreted with care, as their result is dependent on the units of measurement of its arguments (i.e. the value alignment functions).


Obviously, given a value system is defined as a qualitative preference relation between entities, the same value system can be represented by various value system functions.

\begin{definition}[Value System function equivalence]\label{def:vsaf-equivalence-abstract}
Two value system functions $\A_{f_j,{G_V}}$ and $\A_{f_j',{G_V}'}$ are \textbf{equivalent} over a set of entities $E$ if they represent the same value system, i.e. for all $e,e' \in E$:
$$\A_{f_j,{G_V}}(e) \leq \A_{f_j,{G_V}}(e') \iff \A_{f_j',{G_V}'}(e) \leq \A_{f_j',{G_V}'}(e')$$
\end{definition}

Given a specific application domain, we are now in the position to characterize the value system learning problem addressed in this article. It is based on the following assumptions:
\begin{itemize}
    \item A process of \emph{value identification} has been performed \cite{Liscio2023ValuesInSocioTechicalSystemsValueInference} beforehand and a set $V = \{v_1, \dots, v_m\}$ of labels $v_i$ for all relevant values in a particular domain is available. Several value identification approaches from literature can be used for this (e.g.~\cite{wilson2018valueidentificationtext,Liscio2023axiesContextSpecificValueIdentification,QiuZhaoLiLuPengGaoZhu2022ValueNetDataset}).
    \item All agents have a similar understanding of the values in $V$ in the domain at hand, i.e. a consensus grounding function could be found that represents all agents' internal value alignment intuitions to an acceptable degree. Notice that, in general, this need not be the case. For instance, depending on context, the value of fairness can be grounded in preferences that favour ``balanced give \& take'' or ``equal distribution of workload'' \cite{osman2024computationalNUEVO}. However, it is not uncommon for societies that in certain contexts value groundings are effectively socially-agreed upon, e.g. some values in the medical domain~\cite{towardsAwarenessMedicalField2024,userStudyLearningValues} such as autonomy or beneficence. By contrast, value systems are specific to each agent. From a practical point of view, this assumption makes it easier to establish meaningful comparisons among the various value systems learned for different agents, because they are defined over the same value conceptualization. 
    
    \item We have access to experts providing examples of how much an entity is preferred over another with respect to a particular value $v_i$ in a given society. Furthermore, in the specific domain at hand, we can observe examples of rational choices over entities (based on an agent's value system) for each of the agents that we want to assess.
\end{itemize}


The \textit{value system learning problem} addressed in this article consists of solving the following two consecutive tasks: 
\begin{enumerate}
    \item \textit{Value Grounding Learning}: for each value $v_i \in V$, infer an alignment function $\A_{v_i}$ from information about value preferences over entities with respect to $v_i$. These functions constitute the grounding function ${G_V}$ for a particular domain.
    \item \textit{Value System Identification}: based on the grounding function ${G_V}$ and observed examples of rational choices of an agent $j$ (based on its value system $\preccurlyeq^j_{{G_V}}$), infer a value system function $A_{f_j,{G_V}}$ that represents $j$'s value system. According to Definition~\ref{def:value-system-alignment-function} this boils down to learning an adequate aggregation function $f_j$.
\end{enumerate}

Notice that these tasks can be considered independently of each other. For instance, even if the value grounding function was elicited through questionaries, the different value systems could still be learned from observed choices of value-driven rational agents. Furthermore, observe that the value system function learned only needs to be equivalent (not necessarily identical) to the value system function that an agent used to produce the observed behaviour, in the sense of Definition~\ref{def:vsaf-equivalence-abstract}.

In the next subsection, we will instantiate our value system learning problem in the context of sequential decision processes, giving rise to more structured representations and learning tasks.

\subsection{Value system learning in sequential decisions}\label{sec:value-system-learning-in-MDP}

We will now extend the previous definitions and characterise the value learning tasks in the context of sequential decision-making problems. In particular, we would like to learn the value system of an agent based on observations of its behaviour, i.e. in terms of examples of its action sequences (trajectories) in a certain environment.

In particular, we model the environment as Markovian state-transition systems (STS) consisting of a set of states $S$, a set of actions $A$ and a transition function $T: S\times A \times S \to [0,1]$ such that $T(s,a,s')=P(s'|s,a)$ is the probability of reaching state $s'$ if an agent selects action $a$ in state $s$. It should be noted that in our setting the STS not only represents the physical environment but also accounts for the effects of norms and their potential violation~\cite{DBLP:journals/ijar/FagundesOCN16}. For instance, and action $a$ that violates some norm can lead to state $s$ if this violation is detected and punished and to $s'$ otherwise. The non-determinism of the STS accounts for this. In summary, even though our approach does not explicitly represent norms, their potential effects can be considered in the STS world model. 

For a set of trajectories  $\T$  within the STS, the value alignment (see Definition~\ref{def:value-alignment-function}) and value system functions (see Definition~\ref{def:value-system-alignment-function}) are particularized as 
$\A_{v_i}: \T \to \R$ and $\A_{f_j,{G_V}} : \T \to \R$, respectively.

Our running example can also be analysed as a STS where each route option is decomposed into trajectories of states and actions. The STS would be a road network composed by directed road segments (streets/roads) and intersections. The directed segments would be the states of the STS ($S$) and the actions ($A$) at each state would be identified with the next possible segment (state) to transit at the intersection at the end of the current edge. Each state (directed road segment) would be described by two features: the average travel time, and the average fuel consumption of the vehicles that typically transit through it. The transition function $T$ would be deterministic, defined by the existent road network connections.

To analyse these trajectory-based alignment functions further, we make the following assumptions: 
\begin{itemize} 
\item (A1) The alignment of a trajectory $\tau = ((s_0,a_0), \dots, (s_n,a_n))$ with a given value $v_i$ can be calculated as the sum of some reward function over the state-action pairs in $\tau$. In our running example, we can calculate the alignment of a trajectory (road trip) as the sum of the alignment (the \textit{reward}) of the states (road segments) that compose it based on their features.
\item (A2) The value system function that reflects an agent's  preferences over trajectories can be represented by a positive linear combination of the value alignments of the trajectories with regard to the values in $V$. In our running example, the value system aggregation function for agent $j$  ($f_j$) is the linear aggregation function with weights $(2, 1)$ for efficiency and sustainability, respectively.\footnote{Given A2, the set of all possible weight combinations that lead to different value aggregation functions can be described by those in the unit $(m-1)$-simplex ($\Delta^{m-1}$), e.g. the aggregation with weights $(2,1)$ is equivalent to that with weights $(2/3,1/3)\in \Delta^{m-1}$.}.
\item (A3) We assume the human value-based decisions can be \textit{reproduced} by artificial agents that are rational (welfarist utilitarian) value-based decision makers, i.e. their primary criteria for making choices are the alignments of trajectories with each of their values, and their combination based on their value system. Only in case of ties other factors beyond value alignment can be considered.
\end{itemize}    


Assumptions A1 and A2 align conceptually our work with social psychologists like Rohan~\cite{Rohan2000}, who theorized that ``value priorities influence attitudinal and behavioural decisions'', i.e. the selection of trajectories (behaviours) is affected by the importance (or priority) that each person attributes to each value. Despite this, assumption A2 is strict in how this importance is expressed and limits the possible value systems that  agents might hold (and to use in their decision-making). Still, it is still expressive enough to model scenarios where, even though agents judge one value to be more important than another, this can be ``compensated'' by the ``degrees'' of alignment of trajectories with different values. This was seen in the running example for agent $j$ from the previous section: the substantially higher promotion of sustainability of the second alternative compensated its lower efficiency, even though apparently agent $j$ would prefer promoting the latter with its decisions. {In addition, we do not consider agents that prefer to demote certain values (at most, an agent may be indifferent regarding a specific value)}. 

Assumption A1 makes it possible to define the alignments of trajectories with values based on the value alignment of the corresponding transitions. In particular, given a value $v_i \in V$ and a trajectory $\tau = ((s_0, a_0), \dots, (s_n, a_n)) \in \T$ we can define:
\begin{equation}\label{eq:valuereward}
\A_{v_i}(\tau) = \sum_{i=0}^{|\tau|}  R_{v_i}(s_i,a_i),
\end{equation}
where $R_{v_i}: S \times A \to \R$ is some reward function for value $v_i$. 
The function $R_{v_i}$ can be interpreted as a value alignment function over state-action pairs. In fact, given a value alignment function $\A_{v_i}$ for some value $v_i$, for all pairs of trajectories $\tau=((s_0,a_0)), \tau'=((s_0',a_0')) \in \T$ with length $1$ it holds: 

\begin{equation}\label{eq:preferenceontraj}
\tau \preccurlyeq_{v_i} \tau' \iff \A_{v_i}(\tau) \leq \A_{v_i}(\tau') \iff  R_{v_i}(s_0,a_0) \leq R_{v_i}(s_0',a_0')
\end{equation}

Since agents take their decisions in the environment based on their value preferences over trajectories (assumption A3), we can define the STS as a multi-objective Markov Decision Process. We call such processes Markov Value Decision Processes:

\begin{definition}[Markov Value Decision Process]
     A \textbf{Markov Value Decision Process} (MVDP) is a multi-objective Markov Decision Process (MOMDP) represented by a tuple $(S,A,T,V,\Rv_V)$, where:  
     
\begin{enumerate}
    \item $S$ is a set of states.
    \item $A$ is a set of actions agents can take.
    \item $T: S\times A \times S \to [0,1]$ is the transition function, such that $T(s,a,s')=P(s'|s,a)$ is the probability of reaching state $s'$ if an agent selects action $a$ in state $s$.
    \item $V = \{v_1, \dots, v_m\}$ is a finite set of value labels (e.g. $v_1=$`fairness') and upon which agents base their decisions. 
    \item $\Rv_V=\left(R_{v_1}, \dots, R_{v_m}\right)$, is a vector of reward functions such that for all $v_i \in V: R_{v_i}:S \times A \to \R$ is the value reward function for (state, action) pairs with respect to the value $v_i$. We write $\Rv_V(s,a)=\left(R_{v_1}(s,a), \dots, R_{v_m}(s,a) \right)$ to denote the reward vector for a given state-action pair $(s,a)$.
    
    \end{enumerate}

    \end{definition}

For an MVDP $(S,A,T,V,\Rv_V)$ and the set of its feasible trajectories $\T$, due to equation \eqref{eq:valuereward} $\Rv_V$ implicitly defines a grounding function ${G_V}$ of $V$ over $\T$. We say that $\Rv_V$ implements the grounding function ${G_V}$.

With assumption A3, we consider that an agent $j$ will act in the context of an MVDP (whose vector of reward functions $\Rv_V$ implements a grounding function ${G_V}$), by selecting its actions based on its value system $\preccurlyeq_{{G_V}}^j$. It will choose those actions that are part of optimal trajectories, that is, it will try to find the trajectories that are maximally aligned with its value system as per its value system function $A_{f_j,{G_V}}$.

Given a trajectory $\tau = ((s_0, a_0), \dots, (s_n, a_n))$, with Definition~\ref{def:value-system-alignment-function} and equation~\eqref{eq:valuereward} we can rewrite $a$'s value system function as follows:
\begin{equation}\label{eq:vsalignasutility}
\A_{f_j,{G_V}}(\tau)= f_j(\A_{v_1}(\tau), \dots, \A_{v_m}(\tau)) = 
f_j\left(\sum_{i=0}^{|\tau|} R_{v_1}(s_i,a_i), \dots, \sum_{i=0}^{|\tau|} R_{v_m}(s_i,a_i)\right) 
\end{equation}

Considering assumption A2, since $f_j$ consists of a linear combination of its parameters, we can further simplify as follows:
\begin{equation}\label{eq:valuesystem}
\A_{f_j,{G_V}}(\tau)=  
\sum_{i=0}^{|\tau|} f_j(R_{v_1}(s_i,a_i), \dots, R_{v_m}(s_i,a_i)) 
\end{equation}

This means, we can actually transform the calculation of the alignment of a trajectory with an agent's value system to a ``local'' calculation of the transitions that compose the trajectory. In the context of a MOMDP, this allows us to consider $f_j$ as a linear scalarization function of the individual rewards\footnote{A scalarization function in principle, can take any form, being adjustable to very different agent preferences. However, in the context of MOMDPs, a linear scalarization function is often assumed \cite{momdppaperguideHayes2022} because then maximizing the cumulative rewards is equal to maximize the rewards of trajectories. In the literature about inverse reinforcement learning with multiple objectives/rewards the scalarization \cite{moirlMatrixFactorizationKishikawa2021} of objectives via linear weighting is also quite common.}. In this sense, for an agent $j$, its value system function can be considered as a new individual reward function $R_j$ that aggregates the individual value reward functions:
$$R_j(s,a)=f_j(\Rv_V(s,a))$$ 

Thus, if $j$ acts in an MVDP it will employ a policy $\pi_j$ that maximizes the expected accumulated rewards (in practice, considering a maximum horizon $H$):
$$\pi_j(s,a) = \argmax_{\pi}\ \mathbb{E}_{\pi}\left[\sum_{i=0}^{H} R_j(s_i,a_i)|s_0=s,a_0=a\right]$$ 

With this analysis, in the context of learning value systems that drive behaviour in sequential decision-making, the value system learning problem described in the previous subsection can be instantiated as follows:
\begin{enumerate}
    \item \textit{Value Grounding Learning}: for all values $v_i \in V$ learn the (value) reward functions $R_{v_i}$ on state-action pairs from information about the alignment of trajectories with each $v_i$.
    \item \textit{Value System Identification}: 
    based on learned (or given) reward functions $R_{v_i}$ (implementing $G_V$), and observed (rationally) chosen trajectories of an agent $j$ (based on its value system $\preccurlyeq_{G_V}^j$), learn the reward function $R_j$ by aggregating the reward functions $R_{v_i}$ through a scalarization function $f_j$.
    
\end{enumerate}

\section{Algorithms for Value System Learning in MVDP}\label{sec:value-system-learning-algorithms}

In the following, we present algorithms to perform the two aforementioned value system learning tasks in the context of sequential decisions. We consider a MVDP $(S,A,T,V,\Rv_V)$ with $\abs{V}=m$ values. We assume an unknown grounding function ${G_V}$ (implemented by $\Rv_V$) on the values in $V$ that represents the (socially-agreed) semantics of the meanings on the values over trajectories.
Additionally, we consider the existence of a set of agents $J$ with unknown value systems $\preccurlyeq_{{G_V}}^j$ based on ${G_V}$.

We use artificial neural networks for both learning tasks. Like in other deep RL approaches, instead of defining the reward functions on state-action pairs, we consider that they are defined on a space of observable properties of state-action pairs. Thus, we assume some feature space $P$ and a mapping $\phi: S\times A \to P$ that maps state-action pairs to features in $P$. In the sequel we will use either $R_{x}(s,a)$ and $R_{x}(\phi(s,a))$ to refer to the reward for a state-action pair. Furthermore, for simplicity, we extend our definition of reward functions on state-action pairs to trajectories. That is, for any reward function $R_x$ and a trajectory $\tau$, $R_x(\tau) = \sum_{i=0}^{\abs{\tau}} R_x(s_i,a_i)= \sum_{i=0}^{\abs{\tau}} R_x(\phi(s,a))$. This holds for both, value reward functions $R_{v_i}$ with $v_i \in V$ and value system reward functions $R_j$.

Figure \ref{fig:architecture} presents our value system representation model and the instantiation we propose to learn the different parts (value grounding function and value system function) in the sequential decision-making context. As shown, we implement reward functions with neural networks for both the grounding function (to estimate $\Rv_V$) and the value system function (to estimate $R_j$).

\begin{figure}[h]
    \centering
    \includegraphics[width=0.85\linewidth]{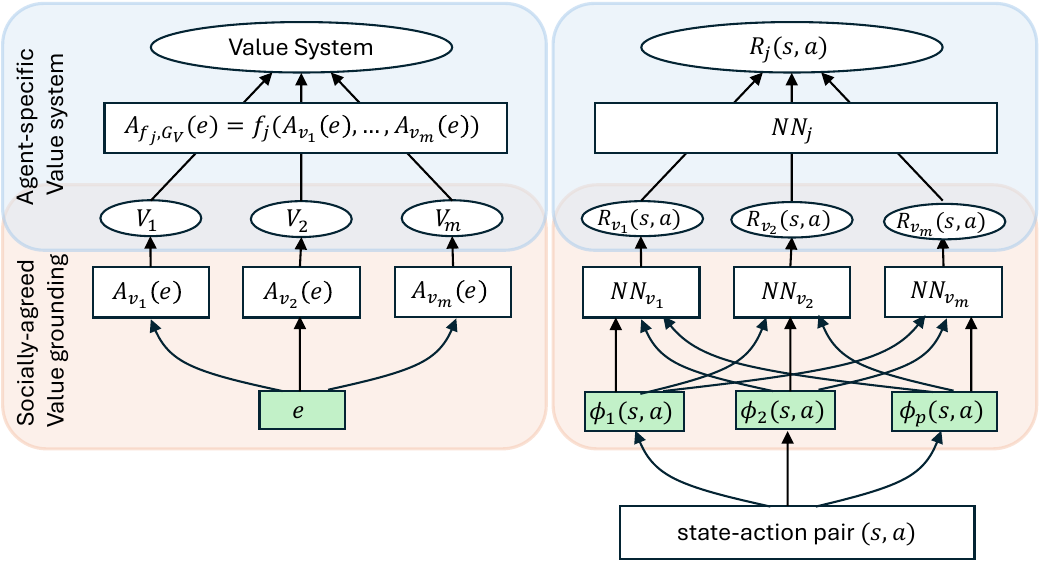}
    
    \caption{Value system representation model (left side) and its instantiation as a value learning problem in sequential decision-making (right side) }
    \label{fig:architecture}
\end{figure}

\subsection{Value Grounding Learning} 
In value grounding learning, we are interested in finding the individual reward functions $R_{v_i}$ for all $v_i \in V$ that form $\Rv_V$ and implement the grounding function ${G_V}$. 

Learning such reward functions can be approached through inverse reinforcement learning (IRL) algorithms~\cite{ng2000algorithms,surveyIRLArora2021}  that learn reward functions from observed optimal behaviour (in this case optimal for value alignment~\cite{andres2024vecompPaper}). However, this approach suffers from some well-known issues: there are many different reward functions that can explain the same behaviour~\cite{ng2000algorithms} and the learned reward functions may not reflect the preferences over suboptimal options. 
For example, consider a simple case with one state $s$ and three actions $a$, $b$ and $c$. Let's suppose that the observed optimal behaviour with regard to some value is always choosing option $a$, because $a$ is preferred over $b$ and also over $c$. Obviously, any reward function that gives a higher reward to $(s,a)$ than to $(s,b)$ or $(s,c)$ would be consistent with the observations. However, such  rewards may not reflect correctly the preference relation between actions $b$ and $c$.

Preference-based inverse reinforcement learning (PbIRL)~\cite{trexpreferences2019} or preference-based reinforcement learning (PbRL)~\cite{surveyRLFromPreferencesPbRL}
are techniques that may provide a way to deal with the aforementioned issue. Some of these methods rely in fitting or learning a reward model~\cite{PreferenceElicitationAndIRLRothkopf2011,furnkranz2012preference,trexpreferences2019} that is close to the ground truth. Rewards are learned from rankings or pairwise comparisons of trajectories instead of just from observed (optimal) trajectories. In this way, the learned reward functions may better estimate an agent's intentions, also upon suboptimal options. However, there may still be different reward functions that represent the same preference relations. In particular, in our context, this approach may not reflect correctly the quantitative difference in the alignment of different alternatives and in the context of our value system representation model this is important. 
We do not need to assure that the learned grounding function is the same as the original grounding function, but we require that it induces the same preference relation over trajectories when the different values are combined in the value system of an agent through the agent's linear aggregation function $f_j$. This is expressed through an equivalence condition between grounding functions(Definition~\ref{def:equivalentgrounding}).

\begin{definition}[Grounding function equivalence]\label{def:equivalentgrounding}
    Let $\Rv_V$ and $\hat{\Rv}_V$ be two reward function vectors that implement the grounding functions $G_V$ and $\hat{G}_V$, respectively.
    $G_V$ and $\hat{G}_V$ are \textbf{equivalent} over the set of trajectories $\mathcal{T}$ under the set of linear aggregation functions $F : \R^m \to \R$ if, for every $f\in F$, and all $\tau,\tau'\in \mathcal{T}$:
    $$\A_{f, G_V} (\tau) \geq \A_{f, G_V} (\tau') \iff \A_{f, \hat{G}_V} (\tau) \geq \A_{f, \hat{G}_V} (\tau')$$
\end{definition}

Since our value system functions are linear combinations of grounding functions (assumption A2), in order to identify correctly the value system of an agent our goal is to learn a grounding function $\hat{G}_V$ that is equivalent to the original $G_V$.

To achieve this goal, in this paper we adapt techniques from PbRL; in particular, our approach is based on~\cite{christiano2023deeprlpreferences}. In this work, a Deep Neural Network approach is presented that learns policies from given qualitative trajectory comparisons, by estimating a reward model. However, our primary focus is on the learned reward functions rather than the resulting policies. Furthermore, to capture the quantitative differences in the alignment of the different options, we incorporate additional information into the learning process. Specifically, we propose providing quantitative comparison pairs of trajectories.

Consider that we want to learn the reward function $\hat{R}_{v_i}$ for value $v_i$ using a neural network structure with parameters $\hat{\theta}\in \Theta$ ($\Theta$ being some parameter space). We write $\hat{R}_{v_i}^{\hat{\theta}}$ to refer to the parametrized reward function. Furthermore, we consider a ground truth reward function $R_{v_i}$ such that $\A_{v_i}(\tau) = R_{v_i}(\tau)$. To learn the full grounding function we will just repeat the next steps for all values in $V$. 

As the input to our learning process we have (for each value $v_i\in V$) a dataset of pairwise comparisons $D_{v_i}$ composed by entries $(\tau, \tau', y)\in D_{v_i}$ where $y \in [0,1]$ indicates quantitatively how much more one trajectory is aligned with value $v_i$ than the other, such that:

\begin{equation}\label{eq:comparisonquantity}
y = \frac{\exp{\A_{v_i}(\tau)}}{\exp{\A_{v_i}(\tau)}+ \exp{\A_{v_i}(\tau')}}
\end{equation}

If $y$ tends to $0$, $\tau'$ is much more aligned with $v_i$ than $\tau$; if it tends to $1$ this is the other way around, and if it is $0.5$ both trajectories are equally preferred with respect to value $v_i$. Equation~\eqref{eq:comparisonquantity} corresponds to the Bradley-Terry model~\cite{bradleyTerryModel1952}, that is used in, e.g.~\cite{christiano2023deeprlpreferences} to approximate comparison preferences in a probabilistic manner. However, as explained, we will interpret this probability model as a deterministic preference quantification method instead.

To estimate $\hat{R}_{v_i}$ we can minimize the following loss over the dataset $D_{v_i}$: 
\begin{equation}\label{eq:groundingloss}
L(\hat{\theta}) =- \frac{1}{\abs{D_{v_i}}}\sum_{(\tau, \tau', y) \in D_{v_i}} y \log\left(p\left(\tau > \tau'|\hat{R}_{v_i}^{\hat{\theta}}\right)\right) + (1-y) \log\left(1-p\left(\tau > \tau'|\hat{R}_{v_i}^{\hat{\theta}}\right)\right)
\end{equation}

Here 
\begin{equation}\label{eq:Bradley}
p\left(\tau > \tau'|\hat{R}_{v_i}^{\hat{\theta}}\right) = \frac{\exp{\hat{R}_{v_i}^{\hat{\theta}}(\tau)}}
{\exp{\hat{R}_{v_i}^{\hat{\theta}}(\tau)} + \exp{\hat{R}_{v_i}^{\hat{\theta}}(\tau')}}
\end{equation}
responds again to the Bradley Terry model. Note that $p\left(\tau > \tau'|\hat{R}_{v_i}^{\hat{\theta}}\right)$ is directly obtained from the current reward estimate $\hat{R}_{v_i}^{\hat{\theta}}$. 

For now, assume that we can find reward functions $\hat{R}_{v_i}$ for all values $v_i \in V$ from sufficient quantitative comparisons (corresponding sets $D_{v_i}$) that represent the seen preferences. Let  $\hat{\Rv}_V$ denote the corresponding reward function vector implementing some grounding function $\hat{G}_V$. 

We can show that the learned grounding function is equivalent to $G_V$ under the set of linear value system aggregation functions, under convergence and some assumptions, with the following two propositions.

\begin{proposition}\label{prop:prop1}
Let $R_{v_i}^{\theta}: S \times A \to \R$ and $\hat{R}_{v_i}^{\hat{\theta}}: S \times A \to \R$ be two reward functions (for value $v_i$) that are parametrized with parameters $\theta,\hat{\theta} \in \Theta$ ($\Theta$ denotes the parameter space). 
Let $D$ a (finite) dataset of preferences with entries $(\tau, \tau', y)$ where:
\[
y = \frac{\exp{R_{v_i}^{\theta}(\tau)}}{
\exp{ R_{v_i}^{\theta}(\tau)} +
\exp{ R_{v_i}^{\theta}(\tau')}
}
\]

Assume that $\hat{\theta}$ minimizes $L$ in equation \ref{eq:groundingloss}, i.e. $\hat{\theta} = \argmin L(\hat{\theta})$. {Additionally, assume that any individual trajectories $\tau, \nu$ appearing in $D$ are directly or indirectly compared, i.e. there is a chain of comparisons that relate $\tau$ and $\nu$.}

Then for all $(\tau, \tau', y)$ $\in D$ it holds:
\[\hat{R}_{v_i}^{\hat{\theta}}(\tau) = R_{v_i}^{\theta}(\tau) + C \text{ and } \hat{R}_{v_i}^{\hat{\theta}}(\tau') = R_{v_i}^{\theta}(\tau') + C\]
and thus
\[ R_{v_i}^{\theta}(\tau) \geq R_{v_i}^{\theta}(\tau') \iff {
\hat{R}_{v_i}^{\hat{\theta}}(\tau) \geq \hat{R}_{v_i}^{\hat{\theta}}(\tau') }
 \]
where $C \in \R$.
\begin{proof}
See Appendix.
\end{proof}
\end{proposition}

This result assures that if a reward function can be represented by a set of parameters $\theta$ (in particular, the ground truth one $R_{v_i}^{\theta} = R_{v_i}$), any other reward function learned (to convergence) with the presented approach, when evaluating on the same trajectories, will only differ from the original reward function by some constant factor. This leads to the next result.

\begin{proposition}\label{prop:prop2}
    Let $V$ be a set of values and $\Rv_V$ and $\hat{\Rv}_V$ be two reward function vectors for $V$ that implement the grounding functions $G_V$ and $\hat{G}_V$, respectively.
    
    If for all trajectories in a set $\mathcal{T}$, $\Rv_V(\tau) = b \cdot \hat{\Rv}_V(\tau) + K$ where $K \in \R^m$ and $b > 0$.  Then, $G_V$ and $\hat{G}_V$ are equivalent in $\mathcal{T}$ under the set of linear aggregation functions.
\begin{proof}
See Appendix.
\end{proof}
\end{proposition}

The above propositions assure that, if we learn reward functions $R_{v_i}$ for all values $v_i \in V$ by minimizing the loss $L(\hat{\theta})$, we find an acceptable grounding function for representing the value system of agents, if those can be represented by linear aggregation functions over the grounding function (as we assume in assumption A2). 

Our approach relies on the availability of a set of pairwise comparisons of example trajectories that quantify the degree of preference $y$ of one trajectory over the other, as defined by equation~\eqref{eq:comparisonquantity}. In some occasions such precise ratings may not be available. In these cases, it may be possible to approximate the values $y$ by more lightweight methods. For example, for each pair $(\tau,\tau')$ of example trajectories, one could ask humans to compare those by rating each trajectory's alignment on some scale $\{1,\dots,X\}$ and then approximate $y$ with:

$$y = \frac{\exp{rate(\tau)}}{\exp{rate(\tau)} + \exp{rate(\tau')}}$$

Notice that it is sufficient that the ratings reflect the comparison between \emph{pairs} of trajectories. In general, rating systems have proven to be useful for reinforcement learning tasks~\cite{ratingRL2024}.~\footnote{Another way of eliciting such ratings could be applied when we are interested in how a society of agents defines the meaning of values. In this context, we may use preference aggregation techniques, which would approximate a consensus $y$ by aggregating binary preferences or ratings from different agents. Indeed, social choice/welfare approaches have been suggested by recent works on PbRL~\cite{proportionalityRLHFAggregationChandakGoelPeters2024}.}

Another factor that influences the applicability of our approach is the required size of the training set of pairwise comparisons over trajectories necessary to properly generalize to unseen trajectories (Proposition~\ref{prop:prop1} does not generalize to trajectories outside of the demonstrations). If the ground truth reward function can be represented with a linear combination over the property space $\phi$, generalization to all trajectories can be guaranteed, in principle, with a relatively small number of comparisons. Proposition \ref{prop:prop3} states that, in the best case, a dataset of comparisons of $p+1$ well-chosen trajectories is sufficient, where $p$ corresponds to the number of features that represent each state-action pair.

\begin{proposition}\label{prop:prop3}
    Let $R_{v_i}^{\theta}$ and $\hat{R}_{v_i}^{\hat{\theta}}$ be two reward functions that are linear in the space of features $\phi: S\times A \to P$, where $P\subset \R^p$.

    Let $\mathcal{U}$  a set of $p+1$ trajectories such that the matrix formed by the cumulative features of each trajectory (extended with a constant column of $1$):

    \begin{equation}
        \Phi = \begin{pmatrix}
        \sum_{(s,a) \in \tau_1} \phi_1(s,a) &\dots &\sum_{(s,a) \in \tau_1} \phi_{p}(s,a) &1\\
        \vdots, &\ddots &\vdots &\vdots\\
        \sum_{(s,a) \in \tau_{p+1}} \phi_1(s,a) &\dots &\sum_{(s,a) \in \tau_{p+1}} \phi_{p}(s,a) &1\\
    \end{pmatrix}
    \end{equation}
    has rank $p + 1$.
    Let $D = \{(\tau, \tau', y)\ |\ \tau, \tau' \in \mathcal{U}\wedge y = \frac{\exp{R_{v_i}^{\theta}(\tau)}}{
\exp{ R_{v_i}^{\theta}(\tau)} +
\exp{ R_{v_i}^{\theta}(\tau')}
} \}$. 

If $\hat{\theta}$ minimizes $L$ in equation \ref{eq:groundingloss},
then $\hat{R}_{v_i}^{\hat{\theta}} = R_{v_i}^{\theta}$.
\begin{proof}
See Appendix.
\end{proof}
\end{proposition}

Of course, in practice it is generally unknown whether the reward vector is actually a linear function of the features and how to choose the comparisons so that the above result is applicable, so datasets should require significantly more than $p+1$ comparisons to correctly reflect the value preferences, as we will see in Section~\ref{sec:eval}.

Algorithm \ref{alg:algorithm1} summarizes the learning algorithm we employ for learning the individual value reward functions $R_{v_i}$ with $v_i \in V$. From a set of quantitative comparisons of different trajectories, the algorithm obtains a hypothesis function $\hat{R}_{v_i}^{\hat{\theta}}$, parametrized with network parameters $\hat{\theta}$, that approximates the function $R_{v_i}$.
  To do that, we perform gradient descent updates (line~\ref{alg:algorithm1-gd}) over the already mentioned PbRL-inspired loss~\cite{christiano2023deeprlpreferences} $L(\hat{\theta})$, in equation~\eqref{eq:groundingloss}, over batches of comparison pairs. The process is repeated for all values in $V$.
\begin{algorithm}[H]
\caption{Value Grounding Learning}\label{alg:algorithm1}
\hspace*{\algorithmicindent} \textbf{Input:} A dataset of preferences
 $D_{v_i}$ (with entries of the kind: $(\tau, \tau', y)$) based on a value $v_i$. A batch size $b$, learning rate $\lambda$ and training steps $N$.\\
\hspace*{\algorithmicindent} \textbf{Output:} a neural network with parameters $\hat{\theta}$ that calculates $\hat{R}_{v_i}^{\hat{\theta}}$.
\begin{algorithmic}[1]
    \State{Initialize network parameters $\hat{\theta}$.}
    \For{training step $t=1,2,\dots, N$}
        \State $B \gets $ choose batch of $b$ tuples from $D_{v_i}$
        \For{each $(\tau, \tau', y) \in B$}
            \State estimate $p\left(\tau > \tau'|\hat{R}_{v_i}^{\hat{\theta}}\right)$ with the current reward estimation $\hat{R}_{v_i}^{\hat{\theta}}$
        \EndFor
        \State\label{alg:algorithm1-gd} $$ \hat{\theta} \gets \hat{\theta} - \frac{\lambda}{b}\frac{\partial}{\partial \hat{\theta}}\left(
\sum_{(\tau, \tau', y) \in B} y \log\left(p\left(\tau > \tau'|\hat{R}_{v_i}^{\hat{\theta}}\right)\right) + (1-y) \log\left(1-p\left(\tau > \tau'|\hat{R}_{v_i}^{\hat{\theta}}\right)\right)\right)$$
    \EndFor
\end{algorithmic}
\end{algorithm}

\subsection{Value System Identification}\label{sec:value-system-identification}
 We will now introduce our approach to learn the value systems of particular agents. We suppose that we have learned de grounding function $G_V$ for a set of values $V$ in the previous step by means of some value reward function vector $\Rv_V$. 
 Given an agent $j$ with unknown value system $\preccurlyeq_{G_V}^j$ based on $G_V$, we want to learn a value system function $\A_{f_j,G_V}$ that represent $\preccurlyeq_{G_V}^j$. 
 
 As we argued above, this actually boils down to learn a new reward function $R_j$ such that $R_j(s,a)=f_j(\Rv_V(s,a))$ where $f_j$ is a linear combination aggregation. Thus, this actually comes down to learn the  weights of the linear combination.
 
For this learning problem we will use traditional inverse reinforcement learning where we have only access to observed samples of an agent $j$'s behaviour, e.g., trajectories. In particular, we estimate a function $\hat{R}_j$ that approximates the real function $R_j$ from a set of training trajectories $\mathcal{T}_j=\{\tau_1,\dots, \tau_\abs{\mathcal{T}_j}\}\subset \mathcal{T}$, supposedly obtained with $j$'s optimal value-base policy $\pi_j$. We use Deep Maximum Entropy IRL~\cite{DeepMaximumEntropyIRLWulfmeier2015}, as it suits well for discrete state-action MDPs. 

Algorithm~\ref{alg:algorithm2} presents the learning procedure. We learn a weight vector $W_j$ that encodes $f_j$ such that $\hat{R}_j(s,a)= W_j \cdot \Rv_V(s,a)$ (dot product) and such that the optimal policy $\hat{\pi}_j$ for $\hat{R}_j$  mimics the original agent policy $\pi_j$. 

The criterion the algorithm uses is the difference in expected state-action visitation counts (i.e. counting state-action visitations instead of solely state visitations as done in the method from~\cite{DeepMaximumEntropyIRLWulfmeier2015}) between the reference policy $\pi_j$ and the learned policy $\hat{\pi}_j$. State-action visitation frequencies over the demonstrations are obtained simply by summing up over the dataset $\mathcal{T}_j$ (line 2). In line \ref{alg:algorithm2-line-sftviteration} we apply the method proposed in \cite{DeepMaximumEntropyIRLWulfmeier2015} to compute the learned policy $\hat{\pi}_j$ from the actual estimations of the reward function $\hat{R}_j$. In line \ref{alg:algorithm2-line-stvisitations}  we employ the technique proposed in \cite{DeepMaximumEntropyIRLWulfmeier2015} to estimate state-action frequencies for the learned policy $\hat{\pi}_j$. In the case we had access to the original policy $\pi_j$ instead of demonstrations ($\mathcal{T}_j$), this method could also be applied to obtain the state-action frequencies $\mu$.

\begin{algorithm}

\caption{Deep Maximum Entropy IRL: Value System Identification}\label{alg:algorithm2}
\hspace*{\algorithmicindent} \textbf{Input:} Let $j$ a particular agent. Set of trajectories 
 $\mathcal{T}_j$ from policy $\pi_j$, reward function vectors $\Rv_V(s,a)$ that implement the grounding function $G_V$, environment horizon $H$, learning rate $\lambda$ and training steps $N$.\\
\hspace*{\algorithmicindent} \textbf{Output:} ${W_j}$ weights that define the function $f_j$
\begin{algorithmic}[1]
    \State{Initialize a weight vector $W_j\in [0,1]^m$.}
    \State $\mu \gets $ \textproc{StateActionVisitationMatrix}($\mathcal{T}_j$)
 
    \For{training step $t=1,2,\dots, N$}
        \State $\hat{R}_j(s,a) = W_j \cdot \Rv_V(\phi(s,a))$

        \State\label{alg:algorithm2-line-sftviteration} $\hat{\pi}_j \gets $ \textproc{SoftValueIterationPolicy}($\hat{R}_j(s,a)$)   
        \State\label{alg:algorithm2-line-stvisitations} $\hat{\mu} \gets $ \textproc{StateActionVisitationMatrix}($\hat{\pi}_j$, horizon=$H$)
        \State $${{W_j}} \gets W_j - \frac{\lambda}{b} \sum_{(s,a) \in S\times A} 
    \left(\hat{\mu}[s,a] - \mu[s,a]\right)\frac{\partial \hat{R}_j(s,a)}{\partial {{W_j}}}$$
    \EndFor
    
\end{algorithmic}

\end{algorithm}

\section{Evaluation}
\label{sec:eval}

We have accomplished simulation experiments in two synthetic use cases to evaluate our learning framework and algorithmic solution approaches: Firefighters and Roadworld. Each environment poses different challenges to the value system problem and algorithms proposed. Replication code of all experiments and environments can be found at \url{https://github.com/andresh26-uam/VAE-ValueLearning/tree/JAAMAS/ValueLearningIRL}.

\subsection{Use cases description}

In this section, we describe the two scenarios used to evaluate our proposal.

\subsubsection{Firefighters}
The Firefighters use case is based on a synthetic MOMDP proposed for a firefighter case simulation \cite{osman2025instillingorganisationalvaluesfirefighters}. It is a simplified two-objective MOMDP environment for an urban high-rise fire scenario, where a firefighter agent must choose among different actions to save people in a building. The objectives encode two values: professionalism ($pf$) and proximity ($px$), i.e. $V = \{pf, px\}$. 

Professionalism refers to acting according to good firefighter practices (e.g. choosing to contain fire only having the adequate equipment, or containing rather than suppressing fire). Proximity refers to acting towards saving the most lives possible, by suppressing fire and evacuating occupants at all costs, even under hard conditions. For example, the action ``evacuate occupants'' is highly aligned with the value of proximity, as it will potentially save people's lives. However, it may not be highly aligned with the value of professionalism, especially when the fire intensity is high and there is little knowledge about the building status, as the action would put the firefighter's own life at risk. The environment consists on $7$ actions and $1200$ possible states with a tabular ground truth reward function $R_{pf}(s,a) \in [-1,1]$, $R_{px}(s,a) \in [-1,1]$. The states are determined on a series of features shown in Table~\ref{tab:state-features-firefighters}. The state transition function and the reward calculations are specified in Table~\ref{tab:transition-firefighters} and Table~\ref{tab:reward-firefighters}, respectively, in the Appendix~\ref{sec:env-specification}.

In this environment, simulated agents can act according to the maximization of either one or the other objective or based on a linear combination of both. According to our model, such a linear combination is the aggregation function of their respective value system function. This MDP has no fixed goal states, rather, firefighters with different value systems will follow a policy that maximizes the cumulative future reward defined by their individual value system function up to a predefined horizon. After some initial experiments, we have found $50$ actions to be a sufficiently large long-term decision horizon. Furthermore, we consider that the agent policies are stochastic\footnote{These stochastic policies are obtained with the soft value iteration procedure in \cite{DeepMaximumEntropyIRLWulfmeier2015}, that assigns a positive probability to each action that is proportional to the exponential future cumulative reward expected upon the action selection.} in order to capture better the dynamics and intrinsic uncertainty of the use case. 

In the grounding learning task we want to approximate the ground truth rewards $R_{pf}(s,a)$, $R_{px}(s,a)$ with a neural network as described in Algorithm~\ref{alg:algorithm1}. The used feature space transformation from state-action pairs $\phi: S\times A \to P$ consists of the $6$ state features (from Table~\ref{tab:state-features-firefighters}) with a finite number of possible values each (each feature value was transformed into its one-hot encoded version to be applicable in the neural network) plus the corresponding action. The candidate reward functions are approximated with a neural network with three fully-connected hidden linear layers (with 50, 100 and 50 neurons each) followed by an output layer with one neuron (with no bias) and a $Tanh$ activation function to get outputs in the reward range $[-1,1]$. Thus, the network computes the reward function $\hat{R}_{v_i}: P \to [-1,1]$. The candidate aggregation functions ($f_j$) for the value system identification task will simply be approximated by a linear neural network layer with no bias term with 2 weights (one per value).

\begin{table}[h!]
    \centering
    \caption{Specification of actions and state features for the Firefighters MDP}
    \label{tab:state-features-firefighters}
    \begin{tabular}{p{0.3\linewidth}|p{0.6\linewidth}}
        \hline
        \textbf{Action} & \textbf{Description} \\
        \hline
        Evacuate Occupants &  Prioritize evacuating people from the building. \\
        Contain Fire &  Focus on containing the fire to prevent it from spreading. \\
        Aggressive Fire Suppression &  Engage in direct firefighting to reduce fire intensity quickly. \\
        Prepare Equipment &  Prepare equipment for safer execution of subsequent tasks. \\
        Update Knowledge & Update the knowledge about the fire and building status to plan subsequent actions. \\
        \hline
    \end{tabular}
    \begin{tabular}{p{0.23\linewidth}|p{0.25\linewidth}|p{0.416\linewidth}}
    \hline
        \textbf{State Feature} & \textbf{Possible Values} &\textbf{Description} \\
        \hline
        Fire Intensity & None, Low, Moderate, High, Severe & Severity of the fire at the current state. \\
        Occupancy & 0, 1, 2, 3, 4 & Area population density. \\
        Equipment Readiness & Not Ready, Ready & Availability of firefighting equipment. \\
        Knowledge & Poor, Good & Building and fire conditions affecting firefighting. \\
        Firefighter Condition & Perfect Health (3), Slightly Injured (2), Moderately Injured (1), Incapacitated (0) & Health status of the firefighter. \\
        \hline
    \end{tabular}
\end{table}

\subsubsection{Roadworld}
In this use case we consider a value-based MDP domain from~\cite{andres2024vecompPaper}, featuring a route choice modelling problem. This domain is adapted from an extract of the Shanghai city road network from~\cite{zhao2023routesairl}. The network is composed of a set of \textit{nodes} that play the role of road intersections and \textit{directed edges} that play the role of streets or road segments that allow the traffic to flow in the specified direction. We assume for simplicity that the capacity of the edges is infinite.

In this network (Figure~\ref{fig:roadworld}), agents move from a certain edge $o$ (the origin) to a predefined destination $d$, using a personal vehicle and selecting the route that aligns most with their particular value preferences.

The states of the corresponding MDP are the set of different edges or road segments (714) and each action consists of choosing the next road segment. The environment in this case is deterministic.

We consider three different values: \textit{sustainability} ($su$), \textit{comfort} ($co$), and \textit{efficiency} ($ef$), i.e. $V = \{su, co, ef\}$. In our simplified scenario, we simulate the corresponding value reward functions ($R_{su}$, $R_{co}$, and $R_{ef}$) directly via specific features of state-action pairs as follows:
\begin{itemize}
    \item $R_{su}(s,a)$: the negative fuel consumption of the road segment given by $(s,a)$.
    \item $R_{co}(s,a)$: a negative cost of the road segment given by $(s,a)$, which is predefined such that road segments with higher travel velocity have a higher comfort cost. 
    \item $R_{ef}(s,a)$: the negative travel time of the road segment given by $(s,a)$.
\end{itemize}
The specific method we used to assign the mentioned features in the network is explained in the Appendix~\ref{sec:env-specification}, Table~\ref{tab:road-segments}. To assure that the previous features are in similar scales, these are normalized by the cost of the maximum cost path (per feature).

Agents acting in the environment are simulated in a way that they choose their routes according to their value system. That is, agents select their routes deterministically maximizing some linear combination of the three negative rewards.

For the grounding learning task of each value we employ a neural network with three inputs, one per each feature describing a state-action pair:  $\phi(s,a) = \left(-fuel(s,a), -comfort\_cost(s,a), -time(s,a) \right)$. A simple linear layer (with no bias term) is used to produce the corresponding reward function $R_{su/co/ef}$. The weights of this network are (before their application over the features), normalized to remain in the range $[0,1]$ and to sum 1 via a \textit{softmax} operation. This, combined with the feature normalization, makes the calculated value alignments remain at similar scales.

The candidate value system functions are, again, approximated by a linear neural network layer, with no bias term and 3 weights (one per value).

\begin{figure}
    \centering
    \includegraphics[trim=14.0cm 8.0cm 12.0cm 8.0cm,clip,width=0.8\textwidth]{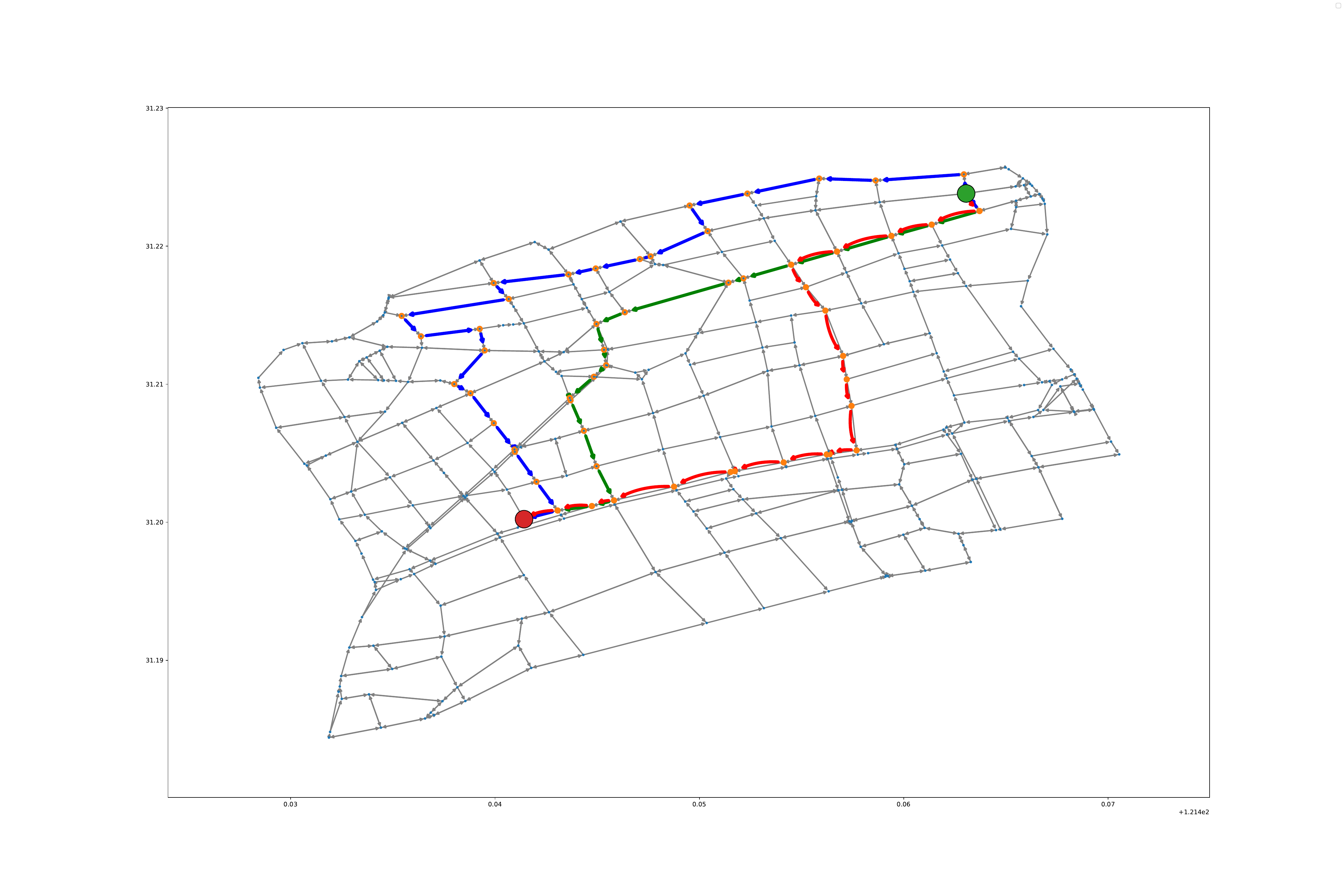}
    \caption{Image depicting the Shanghai road network used for this paper (taken from \cite{zhao2023routesairl}). The green point (upper right part of the figure) is an example origin, and the red point (bottom left) is the destination. The optimal route from the origin to the destination regarding \textit{comfort} is depicted in blue; the optimal route for \textit{sustainability} in green; and the optimal route for  \textit{efficiency} in red.}
    \label{fig:roadworld}
\end{figure}

\subsection{Evaluation: Value Grounding Learning}\label{sec:eval-vgl}
In the first set of experiments we analyse the value grounding learning. As mentioned before, here the objective is to learn reward functions (for each individual value) that maintain the quantitative differences in the value alignment of different compared trajectories with respect to the ground truth differences. To analyse to which extent we can reach this objective, we execute Algorithm~\ref{alg:algorithm1} for all values in both use cases. 

In both use cases, we extract, for each value, a dataset consisting of randomly generated pairs of trajectories (and their quantified comparison) that are sampled from a the policy that maximizes value alignment in a $p$-greedy manner with $p=0.8$ (i.e. that selects, at each state, a value-preferred action by default, or a totally random action with probability $0.8$). 
In the Firefighters domain we sample a total of $10000$ pairs of trajectories (over a pool of 9000 different trajectories). In the Roadworld environment we sample a total of $7000$ comparisons (over a pool of $2500$ trajectories). The trajectories are all of length 50, except that they may be shorter in the Roadworld if they lead to the destination. The trajectory pairs are selected in a manner that we keep a chain of comparisons among all training batches (see Proposition~\ref{prop:prop1}). We accomplish the network training during 200 iterations for each value and all experiments are repeated 10 times with different random seeds.

Figure~\ref{fig:vglrewards} presents the relation between the original (ground truth) rewards and the learned rewards in the corresponding MVDPs. Each dot corresponds to one state-action pair in one of the 10 repetitions. The x-axis represents the original reward value and the y-axis the learned reward value. Thus, if all dots were on the red-dotted line, the learned rewards would be error-free. 

\begin{figure}[ht]
    \centering
    \includegraphics[width=\linewidth]{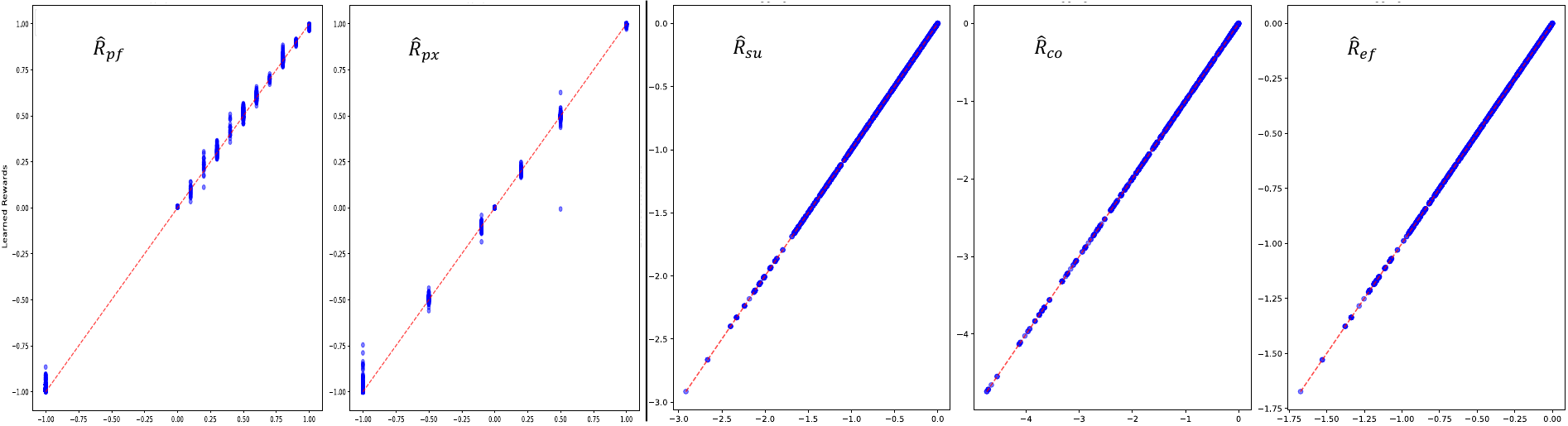}
    
    \caption{Learned value rewards ($\hat{R}_{x}$) versus ground truth rewards ($R_{x}$) for each state-action pair. LEFT: for the two values \textit{professionalism (pf)} and \textit{proximity (px)} in the firefighter environment. RIGHT: for the three values \textit{sustainability (su)}, \textit{comfort (co)} and \textit{efficiency (ef)} in the Roadworld scenario}
    \label{fig:vglrewards}
\end{figure}

As Figure~\ref{fig:vglrewards} shows, in both cases the original rewards are approximated reasonably well. In the Roadworld case, as the original rewards are directly (linearly) related to features, we are able to learn the original functions almost perfectly. In the Firefighters scenario the approximation of the original reward functions is not as good as in the Roadworld case. This is because the reward specification in the environment rewards is generated by context-dependent rules (Appendix~\ref{sec:env-specification}, Table~\ref{tab:reward-firefighters}) that are harder to learn than in the aforementioned Roadworld case. 

It may be noted that in both cases we approximate the real reward functions for all the values in the original ranges. In the Roadworld case this is due to the linearity of the function on the feature vector, as suggested by Proposition~\ref{prop:prop3}. In the case of the Firefighters scenario the reason is that the employed neural network uses a \textit{Tanh} activation on the output neuron which scales the outputs to values in $[-1,1]$ which is exactly the range of the original reward functions.

Albeit the learned reward functions seem to approximate well the original rewards, we want to analyse if they also induce the same preference relations when different rewards are aggregated with the same linear aggregation function. That is, we want to check whether the learned grounding functions are equivalent in the sense of Definition~\ref{def:equivalentgrounding} and thus, will allow us to identify correctly the value systems of different agents in the value system identification task. To accomplish this analysis, we generate a set of 1000 pairs of randomly selected trajectories $(\tau,\tau')$ and check for different linear aggregation functions whether the preference relation between $\tau$ and $\tau'$ is the same for the original grounding function $G_V$ and the learned grounding function $\hat{G}_V$. In particular, for each pair of trajectories, and aggregation function $f$, we evaluate which trajectory is preferred over the other or if they are indifferent with $G_V$, i.e. whether $\A_{f,G_V}(\tau) < \A_{f,G_V}(\tau')$ or $\A_{f,G_V}(\tau) > \A_{f,G_V}(\tau')$ or $\A_{f,G_V}(\tau) = \A_{f,G_V}(\tau')$. Then we check if the relation is correctly maintained when using the learned grounding function $\hat{G}_V$ instead of $G_V$, that is, if 
$\A_{f,\hat{G}_V}(\tau)$ and $\A_{f,\hat{G}_V}(\tau')$ have the same relation. Here, we consider trajectories to be indifferent if their alignment is within a certain tolerance $\epsilon = 0.04$. We consider this epsilon well chosen because the number of comparisons that are affected is below $7\%$ in all cases. The accuracy results are presented in Table~\ref{tab:comparison_use_cases}. As shown, in the Roadworld use case, we obtain a $100\%$ accuracy and in the Firefighters scenario the accuracy is over $98\%$ in all cases. This  indicates that the learned grounding functions induce the original preference relations quite well. 

\begin{table}[ht]
\centering
\begin{minipage}[t]{0.45\linewidth}
\centering
\subcaption{Firefighters Use Case}\label{tab:accuracy_grounding_firefighters}
\begin{tabularx}{\linewidth}{Y|Y}
\hline
$f$ weights & Accuracy [0..1]
\\
\hline
(0.0, 1.0) & 0.987 ± 0.01 \\
(0.2, 0.8) & 0.987 ± 0.01\\
(0.4, 0.6) & 0.992 ± 0.01 \\
(0.6, 0.4) & 0.991 ± 0.01 \\
(0.8, 0.2) & 0.994 ± 0.01 \\
(1.0, 0.0) & 0.995 ± 0.01 \\
\hline
\end{tabularx}
\end{minipage}
\hfill
\begin{minipage}[t]{0.45\linewidth}
\centering
\subcaption{Roadworld Use Case}\label{tab:accuracy_grounding_roadworld}
\begin{tabularx}{\linewidth}{Y|Y}
\hline
$f$ weights & Accuracy [0..1]\\
\hline
(0.0, 0.0, 1.0) & 1.000 ± 0.00 \\
(0.0, 0.33, 0.67) & 1.000 ± 0.00 \\
(0.0, 0.67, 0.33) & 1.000 ± 0.00 \\
(0.0, 1.0, 0.0) & 1.000 ± 0.00 \\
(0.33, 0.0, 0.67) & 1.000 ± 0.00 \\
(0.33, 0.33, 0.33) & 1.000 ± 0.00 \\
(0.33, 0.67, 0.0) & 1.000 ± 0.00 \\
(0.67, 0.0, 0.33) & 1.000 ± 0.00 \\
(0.67, 0.33, 0.0) & 1.000 ± 0.00 \\
(1.0, 0.0, 0.0) & 1.000 ± 0.00 \\
\hline
\end{tabularx}
\end{minipage}
\caption{Preference prediction accuracy for the learned grounding function ($\A_{f,\hat{G}_V}$) for different linear aggregation functions $f$ described by their weights. Results show the average accuracy and sample standard deviation obtained over the 10 experiment repetitions. Table~\ref{tab:accuracy_grounding_firefighters} shows the results in the Firefighters case, Table~\ref{tab:accuracy_grounding_roadworld} shows the results in the Roadworld case}\label{tab:comparison_use_cases}
\end{table}

\subsection{Evaluation: Value System Identification}\label{sec:eval-vsi}

For evaluating the value system identification task, in both domains we analysed how well we are able to identify  the value system aggregation functions (e.g., the linear weights) that represent value systems of different potential agents (Algorithm~\ref{alg:algorithm2}). To do this we simulate agents that apply the following combinations of weights for the sets of values:
\begin{itemize}
    \item Firefighters: $(pf,px)$: $(0,1)$, $(0.2,0.8)$, $(0.4,0.6)$, $(0.6,0.4)$, $(0.8,0.2)$, and $(1,0)$.
    \item Roadworld: $(su,co,ef)$: 
    $(0,0,1)$, $(0,0.33,0.67)$, $(0,0.67,0.33)$, $(0,1,0)$, 
     $(0.33,0.67,0)$, $(0.67,0.33,0)$, $(1,0,0)$,
     $(0.67,0,0.33)$, $(0.33,0,0.67)$, and 
     $(0.33,0.33,0.33)$.
\end{itemize}
Algorithm~\ref{alg:algorithm2} tries to identify the weights by searching for a policy ($\hat{\pi}_j$) that mimics the real policy of an agent ($\pi_j$). Since in our simulation environment we have access to the original policies, instead of training from a set of trajectory demonstrations, we employ the technique to calculate the state-action visitation counts to calculate $\hat{\mu}$ also to calculate $\mu$ in line 2 of the algorithm. In a real world scenario, where there is no direct access to the original policy $\pi_j$, the value of $\hat{\mu}$ has to be estimated from the given example trajectories.

In order to analyse how possible errors in the grounding functions can influence the results of the value system identification task we repeat the experiments for the different weight combinations with both, the original (ground truth) grounding function and the learned grounding function. Each experiment was repeated 10 times and the presented results correspond to averages over the runs.

\subsubsection{Value System Identification in the Firefigthers use case}
Figure~\ref{fig:vsilearningcurve-firefighters} shows the learning curves for the value system identification task in the Firefighters scenario. Specifically, the average total absolute difference in state-action visitation counts (TVC, equation~\eqref{eq:tadvc}) per training iteration is shown to measure as to how far the learned policy approximates the original policy. As it can be seen, the approximation is almost perfect when the original grounding function is used. But notice that even when we use the learned grounding functions (which contain some errors) the approximation to the original policy is still controlled: the final TVC error is less than $0.005$ in all cases.

\begin{equation}\label{eq:tadvc}
    \text {TVC} =\frac{1}{\abs{S\times A}}\sum_{(s,a)\in S\times A}\left|\hat{\mu}[s,a] - \mu[s,a]\right|
\end{equation}

\begin{figure}[H]
    \centering
    \includegraphics[trim=0.0cm 0.0cm 0.0cm 1.75cm,clip,width=0.49\linewidth]{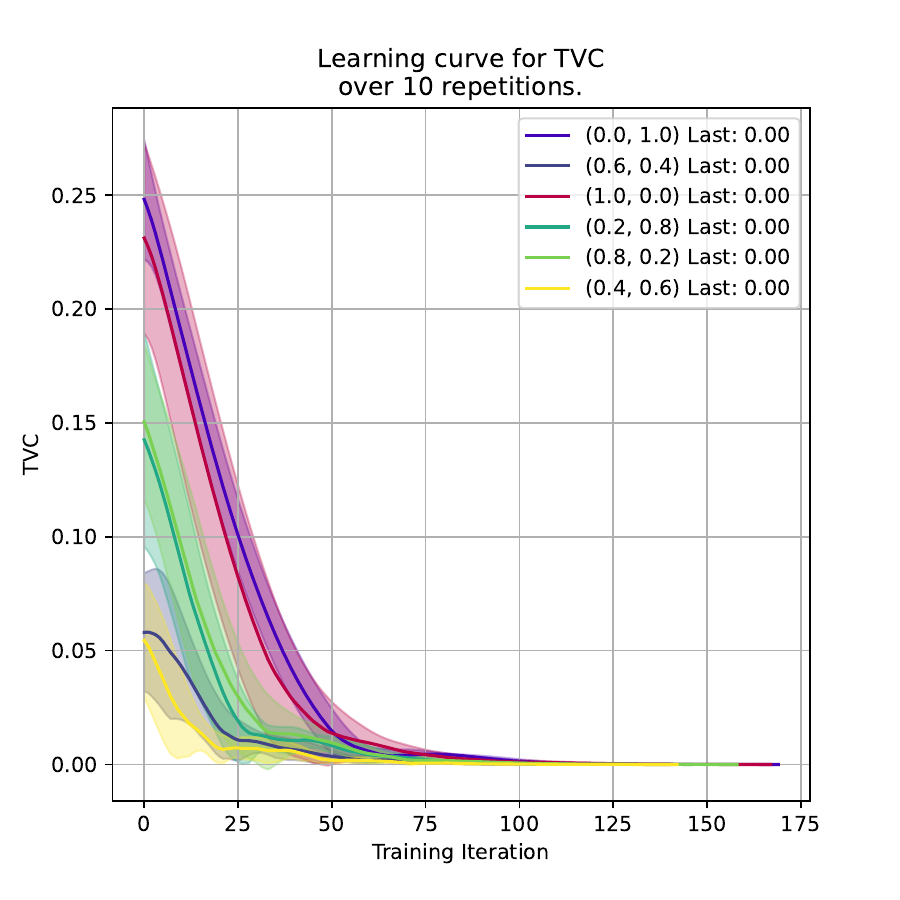}
    \includegraphics[trim=0.0cm 0.0cm 0.0cm 1.75cm,clip,width=0.49\linewidth]{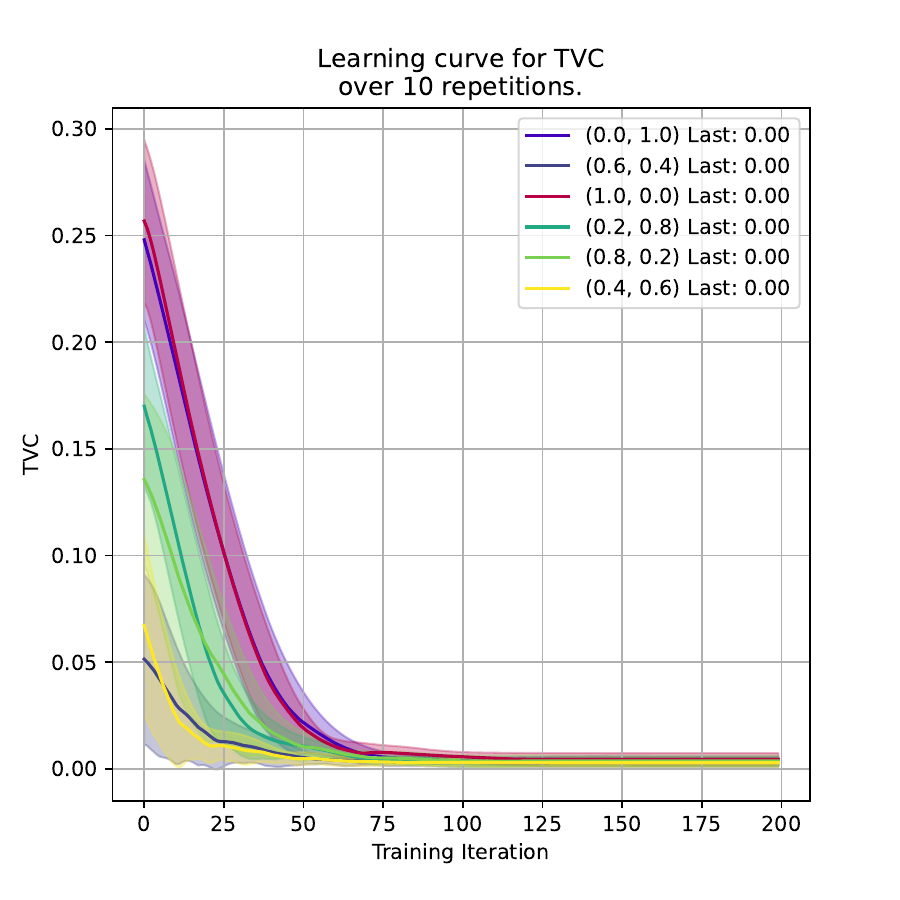}

    \caption{Average total absolute difference in visitation counts (TVC) per training iteration in the Firefighters scenario when performing value system identification assuming the original reward functions $R_{pf}$ and $R_{px}$ (left) and the learned functions (right). In the legend, ``Last'' indicates the TVC obtained in the final training iteration, rounded to 2 decimal places}
    \label{fig:vsilearningcurve-firefighters}
\end{figure}

In order to see whether the learned policies actually lead to trajectories with similar alignments regarding the two values, we generated 1000 sample trajectories with the different original policies  $\pi_j$ and the corresponding learned policies $\hat{\pi}_j$. The results are presented in Tables~\ref{tab:expected_alignments_firefighters_vsi} (with original grounding function) and~\ref{tab:expected_alignments_firefighters_all} (learned grounding function). Shown are the weights of the learned policy and the average (absolute) alignment of the trajectories with \textit{professionalism} ($\A_{pf}$) and \textit{proximity} ($\A_{px}$). The alignment values are calculated over the original (ground truth) alignments, that is, over the original reward functions $R_{pf}$ and $R_{px}$.  

\begin{table}[ht]
\centering
\caption{Firefighters use case. Alignment of sample trajectories with different original ($\pi_j$) and learned policies ($\hat{\pi}_j$) with the two values \textit{professionalism} ($\A_{pf}$) and \textit{proximity} ($\A_{pf}$) when learning from the original grounding function}
\begin{tabularx}{\linewidth}{>{\Centering}p{0.1\linewidth}|Y|Y|Y|Y|Y}

\hline
$\pi_j$ & $\pi_j$: Avg. $\A_{pf}$ & $\pi_j$: Avg. $\A_{px}$ & 
$\hat{\pi}_j$ & $\hat{\pi}_j$: Avg. $\A_{pf}$ & $\hat{\pi}_j$: Avg. $\A_{px}$ \\
\hline
(0.0, 1.0) & 4.262 ±0.035 & 3.509 ±0.036 & (0.000, 1.000) & 4.229 ±0.038 & 3.483 ±0.040 \\
(0.2, 0.8) & 4.398 ±0.039 & 3.417 ±0.022 & (0.200, 0.800) & 4.383 ±0.033 & 3.372 ±0.024 \\
(0.4, 0.6) & 4.532 ±0.033 & 3.275 ±0.035 & (0.400, 0.600) & 4.541 ±0.029 & 3.289 ±0.035 \\
(0.6, 0.4) & 4.639 ±0.030 & 3.154 ±0.051 & (0.600, 0.400) & 4.655 ±0.033 & 3.157 ±0.030 \\
(0.8, 0.2) & 4.777 ±0.056 & 3.045 ±0.042 & (0.800, 0.200) & 4.792 ±0.029 & 3.045 ±0.024 \\
(1.0, 0.0) & 4.891 ±0.062 & 2.910 ±0.053 & (1.000, 0.000) & 4.915 ±0.031 & 2.933 ±0.039 \\
\hline
\end{tabularx}
\label{tab:expected_alignments_firefighters_vsi}
\end{table}

\begin{table}[ht]
\centering
\caption{Firefighters use case. Alignment of sample trajectories with different original and learned policies with the two values \textit{professionalism} ($\A_{pf}$) and \textit{proximity} ($\A_{px}$) when learning from the learned grounding function}
\begin{tabularx}{\linewidth}{>{\Centering}p{0.1\linewidth}|Y|Y|Y|Y|Y}
\hline
$\pi_j$ & $\pi_j$: Avg. $\A_{pf}$ & $\pi_j$: Avg. $\A_{px}$ & 
$\hat{\pi}_j$ & $\hat{\pi}_j$: Avg. $\A_{pf}$ & $\hat{\pi}_j$: Avg. $\A_{px}$ \\
\hline
(0.0, 1.0) & 4.249 ±0.039 & 3.500 ±0.034 & (0.001, 1.000) & 4.241 ±0.043 & 3.481 ±0.040 \\
(0.2, 0.8) & 4.382 ±0.038 & 3.398 ±0.037 & (0.201, 0.800) & 4.391 ±0.027 & 3.377 ±0.034 \\
(0.4, 0.6) & 4.534 ±0.036 & 3.282 ±0.037 & (0.401, 0.601) & 4.523 ±0.042 & 3.261 ±0.047 \\
(0.6, 0.4) & 4.640 ±0.025 & 3.149 ±0.047 & (0.601, 0.401) & 4.660 ±0.029 & 3.159 ±0.027 \\
(0.8, 0.2) & 4.788 ±0.048 & 3.057 ±0.027 & (0.800, 0.201) & 4.809 ±0.029 & 3.055 ±0.034 \\
(1.0, 0.0) & 4.912 ±0.064 & 2.919 ±0.055 & (1.000, 0.002) & 4.910 ±0.025 & 2.918 ±0.031 \\

\hline
\end{tabularx}
\label{tab:expected_alignments_firefighters_all}
\end{table}
Table~\ref{tab:expected_alignments_firefighters_vsi} shows that on the basis of the original grounding function we are able to learn the weights that define the value system of an agent almost perfectly (with exception of some rounding). Furthermore, it is clearly shown that the alignment of the generated trajectories with a value is the higher, the more importance (weight) this value has in the corresponding value system, not only with the original grounding function (Table~\ref{tab:expected_alignments_firefighters_vsi}) but also when the learned grounding function is used (Table~\ref{tab:expected_alignments_firefighters_all}). Also the absolute alignments of the trajectories with the learned policies are very close to the alignments with the original policies. 

Finally, we analysed to which extent the learned value system function is equivalent to the original value system function (e.g., both represent the same preference relation over trajectories, as to Definition~\ref{def:vsaf-equivalence-abstract}). To do that, we use the same idea as in Section~\ref{sec:eval-vgl} and analyse the preference prediction accuracy 
of the learned value system function ($\A_{\hat{f}_j,\hat{G}_V}$) in relation to the original value system function ($\A_{f_j,G_V}$). Again we use a test set of 1000 pairs of randomly selected trajectories $(\tau,\tau')$ and a value of $\epsilon = 0.04$ (also here, the number of comparisons that are affected is below $7\%$).

The accuracy results are presented in Table~\ref{tab:accuracy_firefighters_all}. We only present the results when the learned grounding function is used during the value system identification task. When using the original grounding function instead, the classification accuracy is practically $1$ in all cases, since the learned weight are almost identical to the original weights (as seen in Table~\ref{tab:expected_alignments_firefighters_vsi}). As it is shown, the learned value system functions approximate very well the original value-based preferences of an agent, as the results are all above an accuracy of $0.99$ (more than $99\%$ of correct classifications). It may be noted that even if we only learn from demonstrations from a policy, in this use case the learned function generalizes in a sense that it also evaluates correctly the preferences among sub-optimal trajectories (as seen from the high accuracies over random trajectory comparisons in Table~\ref{tab:accuracy_firefighters_all}).

\begin{table}[ht]
\centering
\caption{Firefighters use case. Preference prediction accuracy for the learned value system functions and the learned grounding function ($\A_{\hat{f}_j,\hat{G}_V}$) for different original value systems ($\A_{f_j,G_V}$).}
\begin{tabularx}{\linewidth}{Y|Y|Y}
\hline
$\A_{f_j,G_V}$ weights & $\A_{\hat{f}_j,\hat{G}_V}$ weights & Accuracy [0..1]\\
\hline

(0.0, 1.0) & (0.001, 1.000) & 0.992 ±0.01  \\
(0.2, 0.8) & (0.201, 0.800) & 0.995 ±0.01 \\
(0.4, 0.6) & (0.401, 0.601) & 0.993 ±0.01 \\
(0.6, 0.4) & (0.601, 0.401) & 0.996 ±0.00 \\
(0.8, 0.2) & (0.800, 0.201) & 0.994 ±0.01 \\
(1.0, 0.0) & (1.000, 0.002) & 0.994 ±0.01 \\

\hline
\end{tabularx}
\label{tab:accuracy_firefighters_all}
\end{table}

\subsubsection{Value System Identification in the Roadworld use case}

Figure~\ref{fig:vsilearningcurve-roadworld} shows the learning curves for the value system identification task in the Roadworld use case. Here we can see that the learned policy approximates the original policy virtually perfectly\footnote{We are certain of this perfect matching because Algorithm~\ref{alg:algorithm2} converged to a $0$ error in state-action visitation counts in all the repetitions of the experiment. This means, the learned policy will choose exactly the same routes to get to the destination from any origin state, except accounting for same cost routes.} in both cases: when learning with the original or with the learned grounding function. This is basically due to the fact that the learned grounding function (the learned value reward functions) is, not only equivalent to the original, as we have shown in Table~\ref{tab:accuracy_grounding_roadworld}, but actually almost identical to the original grounding function (as shown in Figure~\ref{fig:vglrewards}). Thus, the grounding learning task does not introduce additional imprecision in the value system identification task. In fact, this is the case in all experiments we have conducted in this use case. Therefore we only present the results obtained for the value system identification based on the learned grounding function.

\begin{figure}[H]
    \centering
    \includegraphics[trim=0.0cm 0.0cm 0.0cm 1.75cm,clip,width=0.49\linewidth]{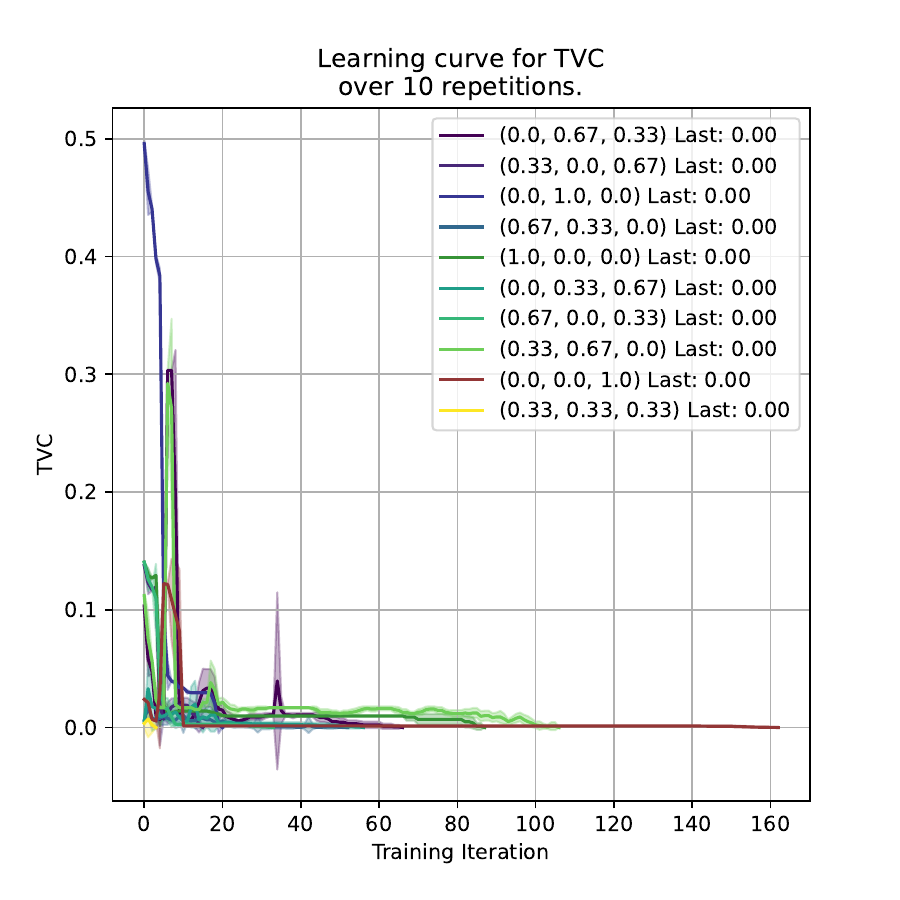}
    \includegraphics[trim=0.0cm 0.0cm 0.0cm 1.75cm,clip,width=0.49\linewidth]{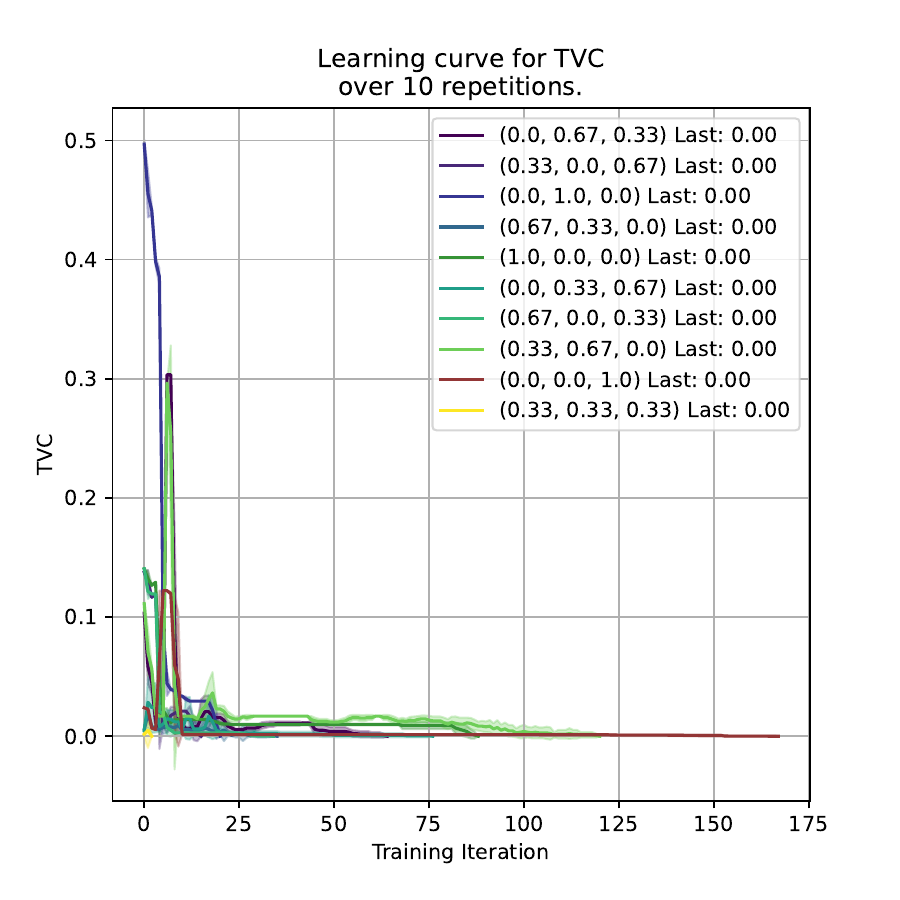}
    \caption{Total absolute difference in visitation counts (TVC) per training iteration in the Roadworld scenario when performing value system identification assuming the original reward functions $R_{su}$, $R_{co}$ and $R_{ef}$ (left) and the learned functions (right). Plotted is the average TVC over 10 repetitions with standard deviations. In the legend, ``Last'' indicates the TVC obtained in the final training iteration, which is exactly $0.0$ in al cases.}
    \label{fig:vsilearningcurve-roadworld}
\end{figure}

Similar to Table~\ref{tab:expected_alignments_firefighters_all} for the Firefighters use case, Table~\ref{tab:expected_alignments_roadworld_all} analyses how much the learned policies actually lead to trajectories with similar alignments with the different values. Again, we set out from 1000 sample trajectories with the different original policies  $\pi_j$ and the corresponding learned policies $\hat{\pi}_j$, present the weights of the learned policy, and compare the average (absolute) alignment of the trajectories with \textit{sustainability} ($\A_{su}$), \textit{comfort} ($\A_{co}$), and \textit{efficiency} ($\A_{ef}$). As the learned policies are totally similar, we expect to have very similar values.

\begin{table}[ht]
\centering
\caption{Alignment of sample trajectories with different original and learned policies with respect to the three values of the Roadworld environment --\textit{sustainability} ($\A_{su}$), \textit{comfort} ($\A_{co}$), and \textit{efficiency} ($\A_{ef}$)-- when learning from the learned grounding function.}\label{tab:expected_alignments_roadworld_all}
\footnotesize
\begin{tabularx}{\linewidth}{>{\Centering}p{0.159\linewidth}|Y|Y|Y|>{\Centering}p{0.191\linewidth}|Y|Y|Y}
\hline
$\pi_j$ & $\pi_j$: Avg. $\A_{su}$ & $\pi_j$: Avg. $\A_{co}$ & $\pi_j$: Avg. $\A_{ef}$ & 
$\hat{\pi}_j$ & $\hat{\pi}_j$: Avg. $\A_{su}$ & $\hat{\pi}_j$: Avg. $\A_{co}$ & $\hat{\pi}_j$: Avg. $\A_{ef}$\\
\hline

(0.0, 0.0, 1.0) & -6.664 $\pm$0.077 & -9.465 $\pm$0.229 & -4.722 $\pm$0.049 & (0.012, 0.011, 0.977) & -6.663 $\pm$0.107 & -9.332 $\pm$0.190 & -4.718 $\pm$0.084 \\
(0.0, 0.33, 0.67) & -6.798 $\pm$0.076 & -8.991 $\pm$0.219 & -4.819 $\pm$0.048 & (0.148, 0.405, 0.447) & -6.795 $\pm$0.107 & -8.858 $\pm$0.218 & -4.817 $\pm$0.083 \\
(0.0, 0.67, 0.33) & -6.978 $\pm$0.070 & -8.340 $\pm$0.168 & -5.047 $\pm$0.044 & (0.127, 0.695, 0.178) & -6.970 $\pm$0.105 & -8.215 $\pm$0.165 & -5.041 $\pm$0.071 \\
(0.0, 1.0, 0.0) & -7.911 $\pm$0.080 & -7.176 $\pm$0.158 & -5.338 $\pm$0.056 & (0.009, 0.982, 0.009) & -7.882 $\pm$0.128 & -7.101 $\pm$0.156 & -5.324 $\pm$0.082 \\
(0.33, 0.67, 0.0) & -6.921 $\pm$0.074 & -8.340 $\pm$0.169 & -5.037 $\pm$0.046 & (0.286, 0.667, 0.047) & -6.921 $\pm$0.104 & -8.237 $\pm$0.166 & -5.029 $\pm$0.069 \\
(0.67, 0.33, 0.0) & -6.766 $\pm$0.078 & -9.061 $\pm$0.224 & -4.813 $\pm$0.047 & (0.541, 0.318, 0.141) & -6.754 $\pm$0.101 & -8.968 $\pm$0.190 & -4.798 $\pm$0.078 \\
(1.0, 0.0, 0.0) & -6.505 $\pm$0.073 & -10.12 $\pm$0.229 & -4.771 $\pm$0.046 & (0.941, 0.004, 0.055) & -6.501 $\pm$0.105 & -10.01 $\pm$0.163 & -4.764 $\pm$0.084 \\
(0.67, 0.0, 0.33) & -6.517 $\pm$0.073 & -10.16 $\pm$0.235 & -4.749 $\pm$0.049 & (0.615, 0.007, 0.377) & -6.513 $\pm$0.104 & -10.04 $\pm$0.169 & -4.746 $\pm$0.082 \\
(0.33, 0.0, 0.67) & -6.556 $\pm$0.073 & -10.08 $\pm$0.234 & -4.726 $\pm$0.048 & (0.409, 0.017, 0.574) & -6.553 $\pm$0.101 & -9.945 $\pm$0.165 & -4.722 $\pm$0.084 \\
(0.33, 0.33, 0.33) & -6.796 $\pm$0.076 & -8.993 $\pm$0.220 & -4.819 $\pm$0.048 & (0.335, 0.351, 0.314) & -6.797 $\pm$0.108 & -8.840 $\pm$0.217 & -4.826 $\pm$0.087 \\

\hline
\end{tabularx}%
\end{table}

As Table~\ref{tab:expected_alignments_roadworld_all} shows, the average alignments of the trajectories obtained from the learned policy (learned value system function) with to the different values are in all cases totally similar to the alignments with the original policy\footnote{Differences are due to the stochasticity in the selection of origin points, and having more than one same minimal cost routes in some cases.}. This assures that the learned policy mimics well the original policy. Notice, however, that in this domain the weights learned for the value system aggregation function ($f_j$) in some cases do not match the original weights. This happens mainly in lines 2, 3, 6 and 9 of the table, and is especially noticeable for the weights learned for  \textit{sustainability} and \textit{efficiency}. The reason for this is the positive correlation among values:  \textit{Sustainability} depends on fuel consumption and \textit{efficiency} on travel time on a road segment. For instance, fuel consumption is lower on roads with higher travel speed (e.g. on primary roads versus on residential areas) and thus, both values will be higher/lower on the same road segments. If such a correlation occurs, our proposed approach can not assure that the original weights are learned, just because different weight combinations (e.g., different preferences over values) may explain exactly the same observations. In fact, as the results show, the weights given to \textit{sustainability} is usually compensated with those given to \textit{efficiency}, and vice versa.
Also, even in the cases where the weight distribution learned is different, as for example in line 2, the average alignment of the trajectories for the learned weights is almost identical to the one obtained with the original weights.  

Finally, like in the Firefighters case, we analysed the preference prediction accuracy of the learned value system function (with learned grounding function) in comparison to the correct preferences relations (represented by the original value system function). We use the same experimental setting as in the Firefighters use case. The results are presented in Table~\ref{tab:accuracy_roadworld_all}.  
Again, we observe that the learned value system functions approximate very well the original value-based preferences of an agent. The prediction accuracy is above $95\%$ in all cases.  Notice that, even though the learning was done from observation of the optimal agent behaviour, in this domain the learned functions generalize well and are able to correctly compare  sub-optimal trajectories.

\begin{table}[ht]
\centering
\caption{Preference prediction accuracy for the learned value system functions and the learned grounding function ($\A_{\hat{f}_j,\hat{G}_V}$) for different original value systems (with original value system function $\A_{f_j,G_V}$) in the Roadworld environment. }
\footnotesize
\begin{tabularx}{\linewidth}{Y|Y|Y} \hline $\A_{f_j,G_V}$ weights & $\A_{\hat{f}_j,\hat{G}_V}$ weights & Accuracy \\ 
\hline 

(0.0, 0.0, 1.0) & (0.012, 0.011, 0.977) & 0.992 $\pm$0.00\\
(0.0, 0.33, 0.67) & (0.148, 0.405, 0.447) & 0.950 $\pm$0.01 \\
(0.0, 0.67, 0.33) & (0.127, 0.695, 0.178) & 0.983 $\pm$0.00 \\
(0.0, 1.0, 0.0) & (0.009, 0.982, 0.009) & 0.998 $\pm$0.00 \\
(0.33, 0.67, 0.0) & (0.286, 0.667, 0.047) & 0.996 $\pm$0.00 \\
(0.67, 0.33, 0.0) & (0.541, 0.318, 0.141) & 0.978 $\pm$0.00 \\
(1.0, 0.0, 0.0) & (0.941, 0.004, 0.055) & 0.990 $\pm$0.00 \\
(0.67, 0.0, 0.33) & (0.615, 0.007, 0.377) & 0.986 $\pm$0.00 \\
(0.33, 0.0, 0.67) & (0.409, 0.017, 0.574) & 0.981 $\pm$0.01 \\
(0.33, 0.33, 0.33) & (0.335, 0.351, 0.314) & 0.983 $\pm$0.02 \\

\hline \end{tabularx}
\label{tab:accuracy_roadworld_all}
\end{table}
\section{Conclusions}\label{sec:conclusion}

With the advance of AI systems in recent years, it has become apparent that in order to
be acceptable to the involved parties,  agreements among intelligent agents must remain aligned with ethical principles and moral values. Value-aware agents extend artificial moral agents~\cite{wallach2008moral} by maintaining an explicit representation of their value system so as to reason with and about it. To permit such reasoning, these artificial agents need a domain-aware representation of  value meaning (we call the value \textit{grounding}) and behave in accordance to the value systems of human stakeholders.

In this article we propose a framework for learning a welfarist utilitarian model of human values and value systems in decision-making environments based on demonstrations of behaviour. This problem, that we call \textit{value system learning}, comprises two tasks. Firstly, value alignment functions for each value relevant in a specific domain must be learned that measure how much an entity is aligned with a certain value. This provides us with the grounding function of a set of values in a particular scenario, which we assume to be feasible to elicit from the possibly diverse (but not incompatible) value understandings of the agents in that domain. Secondly, value system aggregation functions are learned that determine the value system of an agent on top of the grounding function, i.e. a preference relation over the entities present in a specific environment. The goal of the learned model is to serve as an explicit and computable notion of value alignment and of the \emph{pluralistic} value systems in a society that represents observed human value-aligned decisions (even if humans followed other ethical deliberation theories than utilitarianism).

We then apply this framework to the problem of value-aligned sequential decision-making. For this purpose we model the agents' environment as a multi-objective Markov Decision Processes (MOMDP), where the objectives and reward functions directly implement the alignment of the sequential decisions with a given set of values under consideration (their value grounding function). In this framework, value systems (understood  as agent-specific value-based preferences over trajectories) can be represented by a linear scalarization of these rewards. We present a method to learn value grounding functions from quantified pairwise comparisons of trajectories versioning PbRL (Preference-based Reinforcement Learning) approaches, and  show that this method can reliably learn reward functions implementing value grounding functions that are compatible with the original. We also propose the use of inverse reinforcement learning to infer the value systems of agents from observed optimal trajectories, based on the learned rewards. We have implemented algorithms to solve these two tasks, and applied them to perform value system learning for two synthetical MOMDP instances. The value grounding functions and value systems learned  were compatible with the observed behaviour and generalize well to unseen trajectories.

A major lesson learnt from our endeavour is that PbRL techniques are powerful means to learn (quantitative) value alignment functions. They do, however, have the drawback that in real world the size of the training data is limited, as interaction with human experts is required. Furthermore, we conclude that IRL techniques, upon correctly learning the value alignment functions, seem to adequately learn the (ordinal) preference relation implied by linear value system functions. This linear formulation adds interpretability to our results, making explicit the value importance differences observed in the behaviour of one agent and across the full society. In addition, we conjecture that in the real world training data sets are relatively easy to obtain by observing the (optimal) behaviour of agents guided by their value system. 

Still, there are several factors that put limits to the significance of our results. The Firefighter domain has been challenging due to the complex relation between domain features and rewards that had to be learned. By contrast, the Roadworld domain posed the problem of learning value systems with correlated values. However, it is still unclear as to how far our results can be generalised in the context of a value learning problem from real-world data.

A limitation of our framework is the assumption of the observation of an \textit{homogenous} society  where a common value grounding function can be extracted to fairly represent the value understandings of all its members. However, there exist heterogenous societies, e.g. multicultural, where this assumption cannot be considered valid. The proposed value system aggregation functions have another limitation: while in many cases they may be defined as a positive linear transformation of the value grounding function, this need not always be so. Our work is also restricted to  value-based agents whose primary criteria for making choices are the alignments of trajectories with each of their values, and their combination based on their value system. They may only consider other factors, such as individual needs and goals and other motivations~\cite{aguilera2025}, in case of ties -- of course, other types of agent architectures deserve being explored as well. Ultimately, our value system approximation could be useful to help modelling actual human behaviour considered as a module inside broader spectrum theory of mind approaches such as~\cite{theoryofmindpmlr-v80-rabinowitz18a} and agent models used in social simulations~\cite{mercuurJonker2019}.

Regarding future work, our main aim is to identify a real-world domain with sufficient adequate data available to validate our approach under even more challenging conditions. One task on this track is to improve the PbRL approach, and make it less demanding with respect to the required training data. Although we have tested a similar framework that requires only qualitative preferences (instead of the quantitative ones explored in this work) on a real dataset in a non-sequential decision making case~\cite{andresEcai2025}, experiments with real data on sequential decision-making are yet to be carried out. 

Additionally, our approach is limited in expressiveness by the linear (and thus, ``flat'') hierarchy of values used: an interesting avenue to improve this would be integrating our linear value system framework with, e.g. the hierarchical structure of values and intermediate value-related concepts proposed by Osman and D'Inverno~\cite{osman2024computationalNUEVO} (value taxonomies). Another issue of our linear formulation is its limited representativeness of more complex value-based preferences. Hence, non-linear value system aggregation functions, such as using another neural network are worth being explored. However, a neural architecture would seriously harm the interpretability of the current solution. A possible workaround would be investigating context-dependent linear value systems, i.e. considering different value weights under conditions expressed in a interpretable manner. This would maintain interpretability while increasing its capabilities to model complex situations.

The tentative future application of the proposed approach in real-world scenarios raises concerns about possible biases that could be learned from bad quality training datasets. For instance, there could be attributes that are wrongly linked with value meaning due to spurious correlations in the data or biases in the selection of society representatives. To avoid this, datasets for value learning must be collected carefully, ensuring a fair representation of the collective at hand and asking for preference data and trajectories that are relevant and sufficient to describe the task and the environment dynamics. In this line, the static datasets we theoretically analyse here might not be sufficient to completely understand complex environment dynamics; and, for sure, it will not be adaptable to the evolving nature of values~\cite{osman2024computationalNUEVO}. Thus, exploring active learning approaches (e.g.~\cite{pebble2021}) will be necessary in future works. Lastly, in some domains, trajectory and preference data could be classified as sensitive data that needs to be handled with privacy guarantees.

Another line of future work refers to extending our model to the problem of value system learning for societies of agents. An obvious approach to tackle this problem is to learn an aggregation function over the individual agents' value systems using similar techniques to the ones used for (individual) value system aggregation in this paper. However, this would require a cardinal representation of each agent's value system. Notably, in~\cite{andresEcai2025} we have explored this ``social value system learning'' problem in non-sequential decision making, but it is to be tested in sequential scenarios.

Finally, it seems worthwhile to us to explore the relevance of context-dependence for value system learning. In the Roadworld, agents may in general prefer the value of \emph{efficiency}, but when entering densely populated areas, switch to a value system that favours \emph{sustainability}. We will look into how to integrate the notion of context into our model. We also plan to study as to how far context-dependent value systems can be learned efficiently from observations, and whether the ``extent'' of different contexts  can also be inferred automatically. Introducing value systems depending on context might be a way to maintain the interpretability of the proposed linear value system representation while increasing its behaviour and preference modelling capabilities to different conditions (possibly made explicit via rules/decision trees).

\clearpage

\bibliography{sn-bibliography}

\newcounter{lasttable}
\newcounter{lastfigure}
\setcounter{lasttable}{\value{table}}
\setcounter{lastfigure}{\value{figure}}

\begin{appendices}
\renewcommand{\thetable}{\arabic{table}} 
\renewcommand{\thefigure}{\arabic{figure}} 

\setcounter{table}{\value{lasttable}}
\setcounter{figure}{\value{lastfigure}}

\newpage

\section{Glossary of terms}\label{sec:glossary}

\begin{table}[!ht]
    \centering
    \caption{Glossary of terms}\label{tab:glossary-of-terms}
    \begin{tabular}{p{0.37\columnwidth}|p{0.18\columnwidth}|p{0.35\columnwidth}}
    Symbol & Description & Properties \\ 
    \hline
    $\A_{v_i}: E \to \R$    & \textbf{Value alignment function} of value $v_i$  & $\forall e,e' \in E': \A_{v_i}(e) \leq \A_{v_i}(e') \iff e \preccurlyeq_{v_i} e' \text{ w.r.t. value } v_i$. 
    \\
    \hline
    $G_V : E \to \R^m$;  \hspace{1.7cm}\mbox{} $G_V(e)=\left(\A_{v_1}(e),\dots,\A_{v_m}(e)\right)$ & \textbf{Grounding function} of a set of values $V$ &   - \\
    \hline
    Weak order $\preccurlyeq^j_{G_V}$ over $E$ & \textbf{Value system} of an agent $j$ based on grounding function $G_V$ & -  \\
    \hline
    $\A_{f_j,G_V} : E \to \R$ with  
     $\A_{f_j,G_V}(e) = f_j(\A_{v_1}(e), \dots, \A_{v_m}(e))$ & \textbf{Value system function} for agent $j$ & $\forall e,e' \in E: \A_{f_j,G_V}(e) \leq \A_{f_j,G_V}(e') \iff e \preccurlyeq^j_{G_V} e' $. 
     \\
    \hline
    $\A_{f_j,G_V}(e) \leq \A_{f_j,G_V}(e') \iff \A_{f_j',G'_V}'(e) \leq \A_{f_j',G'_V}'(e')$ & Value System function equivalence & - \\
    \hline
    $\tau = ((s_0, a_0), \dots, (s_n, a_n))$ & Trajectory & - \\ 
    \hline
    $R_{v_i}:S \times A \to \R$  & Reward function for value $v_i$ & $\A_{v_i}(\tau) = \sum_{i=0}^{|\tau|}  R_{v_i}(s_i,a_i)$ \\
    \hline
    $(S,A,T,V,\Rv_V)$ & Markov Value Decision Process (MVDP)  &  - \\
    \hline
    $\Rv_V(s,a)=\left(R_{v_1}(s,a), \dots, R_{v_m}(s,a)\right)$ & Reward vector in a MVDP ($m$ values)& Implements a grounding function over trajectories (see below).  \\
    \hline
    $\Rv_V(\tau) = \sum_{i=0}^{|\tau|}  \Rv_V(s_i,a_i)$ & Trajectory grounding (value alignments of $\tau$ with values in $V$)  &  Implements the grounding function in a MVDP, i.e. $R_{v_i}(\tau) =A_{v_i}(\tau)$ \\
    \hline
    $R_j(s,a)=f_j(\Rv_V(s,a))$ & Reward function implementing the value system of agent $j$ with aggregation function $f_j$. & Defines the value system function of $j$ in a MVDP (below).\\
    \hline
    $R_j(\tau)=\sum_{i=0}^{|\tau|} f_j(\Rv_V(s,a))=A_{f_j,G_V}(\tau)$ & Alignment of $\tau$ with the value system of agent $j$ with aggregation function $f_j$. & Value system function of $j$ in a MVDP.\\
    \hline
    $\phi: S\times A \to P$ & Mapping function from state-action pairs to features (properties) of state-action pairs & $R_x(\tau) = \sum_{i=0}^{\abs{\tau}} R_x(s_i,a_i)= \sum_{i=0}^{\abs{\tau}} R_x(\phi(s,a))$  \\
    \hline
    $\hat{G}_V, \hat{R}_{v_i}, \hat{\pi}, ... $   & Learned (estimated) $G_V, R_{v_i}, \pi$, ...  &  \\
    \hline
    \end{tabular}
\end{table}

\clearpage

\section{Proofs of theoretical results}

\begin{proof}[Proof of Proposition~{\upshape\ref{prop:prop1}}]

If the objective $L(\theta)$ is minimized at weights $\hat{\theta}$, then $p(\tau > \tau'|R_{v_i}^{\theta}) = p(\tau > \tau'|R_{v_i}^{\hat{\theta}})$ for all trajectory pairs in $D$. Then, it holds:

    \begin{align}
        p(\tau > \tau'|R_{v_i}^{\theta}) &= p(\tau > \tau'|R_{v_i}^{\hat{\theta}})&\iff\\
        \frac{\exp{R_{v_i}^{\theta}(\tau)}}{
\exp{ R_{v_i}^{\theta}(\tau)} +
\exp{ R_{v_i}^{\theta}(\tau')}} &= \frac{\exp{\hat{R}_{v_i}^{\hat{\theta}}}(\tau)}{
\exp{ \hat{R}_{v_i}^{\hat{\theta}}(\tau)} +
\exp{ \hat{R}_{v_i}^{\hat{\theta}}(\tau')}}&\iff\\
\log\left(\frac{\exp{R_{v_i}^{\theta}(\tau)}}{\exp{\hat{R}_{v_i}^{\hat{\theta}}}(\tau)} \right)&= \log\left(\frac{\exp{ R_{v_i}^{\theta}(\tau)} +
\exp{ R_{v_i}^{\theta}(\tau')}}{
\exp{ \hat{R}_{v_i}^{\hat{\theta}}(\tau)} +
\exp{ \hat{R}_{v_i}^{\hat{\theta}}(\tau')}}\right)&\iff\\
R_{v_i}^{\theta}(\tau)& = \hat{R}_{v_i}^{\hat{\theta}}(\tau) + c(\tau, \tau')
    \end{align}
where 
$$c(\tau, \tau') = c(\tau', \tau)= \log\left(\frac{\exp{ R_{v_i}^{\theta}(\tau)} +
\exp{ R_{v_i}^{\theta}(\tau')}}{
\exp{ \hat{R}_{v_i}^{\hat{\theta}}(\tau)} +
\exp{ \hat{R}_{v_i}^{\hat{\theta}}(\tau')}}\right).$$

Notice that $c(\tau,\tau') = c(\tau',\tau)$, thus,  assuming in the dataset $D$, every pair can be compared to each other, $c$ must actually be a constant $C \in \R$. The argument is as follows:
Starting at $\tau_1$, assume that $(\tau_1, \tau_2, \_) \in D$. From $R_{v_i}^{\theta}(\tau_1) = \hat{R}_{v_i}^{\hat{\theta}}(\tau_1) + c(\tau_1, \tau_2)$ and $R_{v_i}^{\theta}(\tau_2)= \hat{R}_{v_i}^{\hat{\theta}}(\tau_2) + c(\tau_2, \tau_1)$ we name a $C = c(\tau_2, \tau_1) = c(\tau_1, \tau_2)$. 

By induction, assume, that for all $n\leq N$:
$$R_{v_i}^{\theta}(\tau_n) = \hat{R}_{v_i}^{\hat{\theta}}(\tau_n) + C$$
where $C =  c(\tau_{N-1}, \tau_{N}) = \dots =  c(\tau_1, \tau_2)$
Let  $(\tau_{N}, \tau_{N+1}, \_) \in D$. This implies $R_{v_i}^{\theta}(\tau_{N})= \hat{R}_{v_i}^{\hat{\theta}}(\tau_{N}) + c(\tau_{N}, \tau_{N+1})$. But we had $R_{v_i}^{\theta}(\tau_N)= \hat{R}_{v_i}^{\hat{\theta}}(\tau_N) + C$, then $c(\tau_N, \tau_{N+1}) = C$. By induction, as the chain covers all pairs in $D$,
we have, $R_{v_i}^{\theta}(\tau) = \hat{R}_{v_i}^{\hat{\theta}}(\tau) + C$ for all $\tau$ among all trajectories present in $D$.

Thus, $R_{v_i}^{\theta}(\tau) > R_{v_i}^{\theta}(\tau') \iff \hat{R}_{v_i}^{\hat{\theta}}(\tau) > \hat{R}_{v_i}^{\hat{\theta}}(\tau')$, for all $(\tau,\tau')\in D$.

\end{proof}

\begin{proof}[Proof of Proposition~{\upshape\ref{prop:prop2}}]
    Let $f$ be a linear aggregation function with weight vector $(w_1,\dots , w_m)$ and let $\tau, \tau' \in \T$ be any two trajectories. Furthermore, assume $ \hat{\Rv}_V(\tau) = b \cdot \Rv_V(\tau) + K$.
    
    It holds: 
    \begin{align}   
        \A_{f,{G_V}} (\tau) \geq \A_{f,{G_V}} (\tau')  \iff \\
        f \cdot \Rv_V(\tau) \geq f \cdot \Rv_V(\tau')  \iff \\ 
        f \cdot b \cdot \Rv_V(\tau) \geq f \cdot b \cdot \Rv_V(\tau') \iff \\ 
        f \cdot b \cdot \Rv_V(\tau) + f \cdot K \geq f \cdot b \cdot \Rv_V(\tau') + f \cdot K \iff \\ 
        f \cdot (b \cdot \Rv_V(\tau) + K) \geq f \cdot (b \cdot \Rv_V(\tau') + K) \iff \\ 
        f \cdot \hat{\Rv}_V(\tau)  \geq f \cdot \hat{\Rv}_V(\tau') \iff \\ 
        \A_{f,\hat{G}_V} (\tau) \geq \A_{f,\hat{G}_V} (\tau')  
    \end{align}
    which proofs the proposition. 
\end{proof}

\begin{proof}[Proof of Proposition~{\upshape\ref{prop:prop3}}]

    Under the conditions of the proposition, from the minimization of the loss \eqref{eq:groundingloss}, it follows that $p(\tau > \tau'|R_{v_i}^{\theta}) = p(\tau > \tau'|R_{v_i}^{\hat{\theta}})$ for all trajectories in $D$. 

With the assumption that $D$ consists in all comparisons between trajectories from $\mathcal{U}$, there must be a chain of comparisons covering all $D$. We can apply Proposition~\ref{prop:prop1} to see that there exists $K\in \R$ such that $R_{v_i}^{\theta}(\tau) = \hat{R}_{v_i}^{\hat{\theta}}(\tau) + K$ for all $\tau \in \mathcal{U}$. 

Now, we use that $R_{v_i}^{\theta}(\tau)$ and $\hat{R}_{v_i}^{\hat{\theta}}(\tau)$ are linear in the space of properties; then, we can write:
\begin{align*}
    R_{v_i}^{\theta}(\tau) &= \sum_{(s,a) \in \tau} R_{v_i}^{\theta}(s,a)
    &= \sum_{(s,a) \in \tau} \begin{pmatrix}
        \theta_1\\
        \vdots\\
        \theta_p
    \end{pmatrix}\phi(s,a)
    &= \begin{pmatrix}
        \theta_1\\
        \vdots\\
        \theta_p
    \end{pmatrix} \sum_{(s,a) \in \tau} \phi(s,a)
\end{align*}
\begin{align*}
    \hat{R}_{v_i}^{\hat{\theta}}(\tau) &= \sum_{(s,a) \in \tau} \hat{R}_{v_i}^{\hat{\theta}}(s,a)
    &= \sum_{(s,a) \in \tau} \begin{pmatrix}
        \hat{\theta}_1\\
        \vdots\\
        \hat{\theta}_p
    \end{pmatrix}\phi(s,a)
    &= \begin{pmatrix}
        \hat{\theta}_1\\
        \vdots\\
        \hat{\theta}_p
    \end{pmatrix} \sum_{(s,a) \in \tau} \phi(s,a)
\end{align*}

From the results of Proposition~\ref{prop:prop2}, we can write the following compatible system of equations (which variables are the difference in the weights of the reward functions):
\begin{equation}\label{eq:system_prop3}
    \Phi' \cdot \left(\begin{pmatrix}
        \theta_1\\\vdots\\\theta_p
    \end{pmatrix} - \begin{pmatrix}
        \hat{\theta}_1\\\vdots\\ \hat{\theta}_p
    \end{pmatrix}\right) = \begin{pmatrix}
        K\\\vdots\\K
    \end{pmatrix}
\end{equation}

where 
\begin{equation*}
        \Phi' = \begin{pmatrix}
         \sum_{(s,a) \in \tau_1}  \phi_1(s, a) &\dots &\sum_{(s,a) \in \tau_1}  \phi_p(s, a)\\
        \vdots, &\ddots &\vdots \\
        \sum_{(s,a) \in \tau_{p+1}}  \phi_{1}(s, a) &\dots &\sum_{(s,a) \in \tau_{p+1}}  \phi_p(s, a)\\
    \end{pmatrix}
    \end{equation*}

Assume that $K \neq 0$. As $\Phi$ has rank $p+1$, the extended matrix of \eqref{eq:system_prop3} has also rank $p+1$:

\begin{equation*}
        \Phi'' = \begin{pmatrix}
        \sum_{(s,a) \in \tau_1}  \phi_1(s, a) &\dots &\sum_{(s,a) \in \tau_1}  \phi_p(s, a) & K\\
        \vdots, &\ddots &\vdots & \vdots \\
        \sum_{(s,a) \in \tau_{p+1}}  \phi_{1}(s, a) &\dots &\sum_{(s,a) \in \tau_{p+1}}  \phi_p(s, a) & K\\
    \end{pmatrix}
    \end{equation*}
But then, by Rouch{\'e} Frobenius, the system~\eqref{eq:system_prop3}  has no solution (which contradicts the fact that it is compatible). Therefore, it must hold that $K = 0$.

Then, $\Phi''$ has rank $p$, thus, system~\eqref{eq:system_prop3} has a single solution, which must given by 
$$\begin{pmatrix}
        \theta_1\\\vdots\\\theta_p
    \end{pmatrix} - \begin{pmatrix}
        \hat{\theta}_1\\\vdots\\ \hat{\theta}_p
    \end{pmatrix} = \begin{pmatrix}
        0\\\vdots\\0
    \end{pmatrix}$$

It follows that $R_{v_i}^{\theta} = \hat{R}_{v_i}^{\hat{\theta}}$ for all $S\times A$.
\end{proof}

\section{Specification details of the evaluation environments}\label{sec:env-specification}

\subsection{Firefighters environment}
In Table~\ref{tab:reward-firefighters} we specify the ground truth reward in the Firefighters MDP for each of the two values, given the current state, action and the next state. The next state is not strictly needed as information to be able to infer this reward function, as it is calculated deterministically with the transition rule in Table~\ref{tab:transition-firefighters}.

\begin{table}[h]
    \centering
    \caption{Firefighter Domain State-Transition Function. Legend: KN = Visibility, FI = Fire Intensity, EQ = Equipment Readiness, OC = Occupancy, FFC = Firefighter Condition, FL = Floor Level}
    \label{tab:transition-firefighters}
    \begin{tabular}{|l|l|l|}
        \hline
        \textbf{Action} & \textbf{State Conditions} & \textbf{State Transitions} \\
        \hline
        \multirow{3}{*}{Evacuate Occupants} 
            & Always & OC $\rightarrow \max(0, \text{OC} - 1)$ \\
            & FI $\geq 3$, EQ = 0, KN = 0 & FFC $\rightarrow \max(0, \text{FFC} - 1)$ \\
            & FI = 5 & EQ $\rightarrow 0$ \\
        \hline
        Contain Fire & Always & FI $\rightarrow \max(0, \text{FI} - 1)$ \\
        \hline
        \multirow{3}{*}{Aggressive Fire Suppression} 
            & Always & FI $\rightarrow \max(0, \text{FI} - 2)$ \\
            & FI $\geq 3$, (EQ = 0 $\vee$  KN = 0) & FFC $\rightarrow \max(0, \text{FFC} - 1)$ \\
            & FI = 5 & EQ $\rightarrow 0$ \\
        \hline
        Prepare Equipment & Always & EQ $\rightarrow 1$ \\
        \hline
        Update Knowledge & Always & KN $\rightarrow 1$ \\
        \hline
    \end{tabular}
\end{table}

\begin{table}[h]
    \centering
    \caption{Firefighter Domain Reward Specification. Legend: KN = Knowledge, FI = Fire Intensity, EQ = Equipment Readiness, OC = Occupancy, FFC = Firefighter Condition. Movement actions (``Go Upstairs/Go Downstairs'') by default yield a reward of $0$ for all values. To note is that if the firefighter is incapacitated in the next state, the rewards are $-1.0$ overriding any other situation. }\label{tab:reward-firefighters}
    \begin{tabular}{|l|l|c|c|}
        \hline
        \textbf{Action} & \textbf{State Conditions} & \textbf{Professionalism} & \textbf{Proximity} \\
        \hline
        \multirow{2}{*}{Evacuate Occupants} 
            & OC = 0 & -1.0 & -1.0 \\
            & OC $\neq$ 0 & $1 - 0.2 \times \text{FI} - 0.1 \times \text{KN}$ & 1.0 \\
        \hline
        \multirow{2}{*}{Contain Fire} 
            & FI = 0 & -1.0 & -1.0 \\
            & FI $\neq$ 0 & 0.8 & 0.2 \\
        \hline
        \multirow{3}{*}{Aggressive Fire Suppression} 
            & FI = 0 & -1.0 & -1.0 \\
            & FI $\neq$ 0, EQ = 0 & 0.3 & 0.5 \\
            & FI $\neq$ 0, EQ $\neq$ 0 & 0.6 & 0.5 \\
        \hline
        \multirow{2}{*}{Prepare Equipment} 
            & EQ = 0 & 0.5 & -0.1 \\
            & EQ $\neq$ 0 & -1.0 & -1.0 \\
        \hline
        \multirow{2}{*}{Update Knowledge} 
            & KN = 0 & 1.0 & -0.5 \\
            & KN $\neq$ 0 & -1.0 & -1.0 \\
        \hline
        \multicolumn{2}{|c|}{Next state has FFC = 0 (Incapacitated)} & -1.0 & -1.0 \\
        \hline
    \end{tabular}
\end{table}

\subsection{Roadworld environment}

In the Roadworld domain, the network road segments have assigned a set of 3 costs that act as features and upon which rewards are obtained: $fuel$ (estimated fuel consumption), $comfort\_cost$ (discomfort), and $time$ (expected travel time).

We use a simple rule to obtain these features: the costs are based on fixed weights per road type multiplied by the length of the road segment. 
The data is from OpenStreetMap, taken from the dataset in \cite{zhao2023routesairl} and the weights for each road type are summarized in Table~\ref{tab:road-segments}.

\begin{table}[htbp]
    \centering
    \caption{Cost weights for $fuel$, $comfort\_cost$, and $time$ applied in the Roadworld scenario.
    }
    \label{tab:road-segments}
    \begin{tabularx}{\linewidth}{|X|Y|Y|Y|}
        \hline
        \textbf{Road Type} & \textbf{fuel} & \textbf{comfort\_cost} & \textbf{time  
        } \\
        \hline
        Residential & 20.0 & 1.0 & 66.67 \\
        Primary     & 12.0 & 30.0 & 14.29 \\
        Unclassified & 20.0 & 1.0 & 25.0 \\
        Tertiary    & 7.0  & 8.0 & 50.0 \\
        Living Street & 25.0 & 1.0 & 66.67 \\
        Secondary   & 9.0  & 15.0 & 50.0 \\
        \hline
    \end{tabularx}
\end{table}

\end{appendices}
\end{document}